%% file: MPIGH_fR2body_VerEPJC.tex
\documentclass[preprint,floatfix,showpacs,footinbib,amsmath,amssymb,amsfonts,superscriptaddress,verbatim,latexsym,minimal]{revtex4-2}
\usepackage{bm}
\usepackage{graphicx}
\usepackage{color}
\usepackage{dsfont}
\usepackage[normalem]{ulem}
\usepackage{hyperref}
\usepackage{mathtools}
\usepackage{mathrsfs}
\usepackage{calrsfs}
\usepackage{comment}
\usepackage[utf8]{inputenc}
\usepackage{varioref}
\usepackage[active]{srcltx}

\expandafter\ifx\csname package@font\endcsname\relax\else
\expandafter\expandafter
\expandafter\usepackage
\expandafter\expandafter
\expandafter{\csname package@font\endcsname}%
\fi
\hypersetup{
	colorlinks=true,       
	linkcolor=blue,          
	citecolor=blue,        
}

\usepackage[section]{placeins}

\usepackage{fancybox}
\labelformat{figure}{Fig.~#1} 

\DeclarePairedDelimiter\floor{\lfloor}{\rfloor}

\begin{document}
	\title{The Equivalence and/or the Effacing principle in $ f\left(R\right) $ theories of gravity} 
	\author{Soham Bhattacharyya}
	\affiliation{Max Planck Institute for Gravitational Physics (Albert Einstein Institute)\\ Leibniz Universit\"at Hannover}
	\email{soham.bhattacharyya@aei.mpg.de}
	\begin{abstract}
		\noindent
		The Einstein-Hilbert action of the general theory of relativity (GR) is the integral of the scalar curvature $R$. It is a theory that is drawn from the Equivalence principle and has predictions that come out as a consequence of the principle, in observables. Testing such observables to find confirmation/infirmation of the principle has formed a significant chunk of tests of GR itself. It is expected that quantum corrections to GR may add additional higher powers of $R$ to the Einstein-Hilbert action, or more generally, modifying the action into a generic class of functions of the Ricci scalar. Testing the fate of the prized equivalence principle, in such modified theories of gravity, hence become important in order to obtain a more generic theory of gravitation, and consequently, of gravitating objects. In this study, it is shown that a Post-Newtonian (PN) expansion of a class of $ f\left(R\right) $ theories lead to a sequence of solutions to the two-body problem, which follows the equivalence principle (EP) at the Newtonian order, and generalizes to the 'effacing principle' at a higher PN order.
	\end{abstract}
	\maketitle
	\section{Introduction}
	\label{s:intro}
	The equivalence principle remains to this date a cornerstone in an attempt to find the most general theory of gravitation possible. Testing the principle of equivalence has almost become synonymous with testing GR at all possible length scales, as illustrated in \cite{Will:2014kxa}. However, it is entirely possible that the picture is more subtle than that.\\
	
	The Equivalence principle (EP), being a principle, is not a theory, and its consolidation from assumption to fact in the scientific community is based on empirical (but possibly circumstantial) evidence. Hence it becomes all the more important to keep on trying to find exceptions to the rule in natural phenomenon to test the limits of a scientific principle. The physical consequences of EP are found through the predicted effects of Galilean, Newtonian, and Einsteinian relativity. Namely, the trajectory of an extended object in a space-time with other gravitational sources, is independent of the internal gravitational structure of the extended body, till at a scenario where the ratio of object extendedness to inter-object distances become significant. Tidal deformations of individual objects then strongly affect the gravitational field of the external space-time, and structural details of individual objects begin to play a significantly important role in the evolutionary dynamics of the equations of motion of such compact objects.\\ 
	
	Various tests of the equivalence principle have been attempted, some of which date quite far back into the past, like the case of John Philoponus in the 6th century \cite{2014philoponus}, Galileo's tests in 1610 \cite{galilei2013discorsi}, and Newton's pendulum experiments in 1680 \cite{newton1871philosophiae}. However, the most rigorous tests of the equivalence principle in the pre-modern times were the torsion pendulum experiments of E\"otv\"os in \cite{eotvos1907bestimmung} and following publications. Modern tests of the EP include \cite{ROLL1964442,Schlamminger2008}, whereas strong field tests of the strong EP was performed in \cite{2014Natur.505..520R,2018Natur.559...73A}\\
	
	The most extreme test of the equivalence principle conceived so far is through the analysis of gravitational wave (GW) data. In GW physics, the test of the equivalence principle, or a consequence of it, is known as the 'effacing principle' in literature \cite{Damour:1982wm,Damour:1984rbx}. The principle can be stated as follows: the internal gravitational (strong or weak) details of  extended compact objects in a binary system, neutron stars (NS) or black holes (BH), does not influence the eventual evolution of the trajectory of the individual objects, till at a very later stage of the orbital evolution. In other words, the deformation of a compact object due to the presence of another compact object in its vicinity, although changing the structure of the first object, does not influence the gravitational field of the external space-time, and the dynamics of the center of mass (COM) of the massive compact objects remain unaffected by said deformations. The effacing principle is broken at scenarios where the average radii of the objects become comparable to the distances between objects where tidal deformations have significant effects on the orbital dynamics. Information about the equations of motion of a binary system, for example, comes to an observer at asymptotic infinity through the phase evolution of GWs. Studying such data, one can estimate the efficacy of GR in terms of obtaining an accurate relationship between orbital dynamics and GW data.\\
	
	In terms of testing the principle, if objects move on geodesics of an external gravitational field, then the rate at which they rotate and fall around (towards) each other is fixed only by their masses and orbital separation. Perturbatively solving GR field equations from a Newtonian/Keplerian initial data leads to a sequence of solutions that obey the effacing principle till 2.5 Post Newtonian order \cite{Damour:1982wm}, for two non-spinning or slowly spinning compact objects in a binary system. That is, their equations of motion are dependent only on their masses and distance between their center of masses, till at a much later stage of the orbital evolution.\\
	
	However, consequences of the effacing principle might not be unique to GR. To show this, we take a class of theories which are generalizations of the Einstein-Hilbert action of GR, collectively called as $ f\left(R\right) $ theories of gravity \cite{Starobinsky2007,Nojiri:2006ri}, and expand an auxiliary field about a Minkowski space-time using the Post-Newtonian formalism. The action $ S $ from which the field equations of $ f\left(R\right) $ theories of gravity are obtained is given by the following 4-integral over a manifold $ \mathcal{M} $
	\begin{eqnarray}\label{fRaction}
	S &=& \int_{\mathcal{M}}\,\sqrt{-g}\,\,d^4x\,\left[\frac{f\left(R\right)}{2\,\kappa^2}\,+\,T_m\right]
	\end{eqnarray}
	where $ f\left(R\right) $ is a generic function of the Ricci scalar $ R $, and $ \sqrt{-g}\,d^4x $ is the covariant infinitesimal volume element. $ T_m $ is the classical matter action. The above reduces to the familiar Einstein-Hilbert action for $ f\left(R\right)\,=\,R $, and the field equations of GR can hence be recovered. Geometrized units, $ c\,=\,G\,=\,1 $ will be used in this study, which implies: $ \kappa^2\,=\,8\,\pi $. The notations of \cite{Futamase2007} will be followed except a few changes in variable and index labeling.\\
	
	In this article, the two-body problem in $ f\left(R\right) $ theories of gravity will be solved for, using an initial data on the metric and matter that is asymptotically Newtonian/Keplerian A relaxed form of the field equations will be used, similar to the Landau-Lifshitz formulation of GR in \cite{LANDAU1975345,*LANDAU1975295,*LANDAU1975259}, leading to a post-Newtonian sequence of solutions, as reviewed in Sec. \ref{s:EqEfGR}. The solutions will be expressed as functions of multipole moments over two compact sources, and it will be explicitly shown that the equations of motion of the two sources are dependent only on the monopole moments of the sources, and the distance between their 'center of masses'.\\
	
	In Sec. \ref{s:EqEffR} a method to relax the field equations of $ f\left(R\right) $ theories will be illustrated. Using the formalism of Sec. \ref{s:EqEfGR} "gauge invariant" modification to the 6 PN equations of motion of GR due to $ f\left(R\right) $ will be derived. It will also be shown that even though deviations from general relativistic equations of motion exist, consequences of the equivalence (and consequently the effacing) principle appear in the equations of motion of two compact objects in $ f\left(R\right) $ theories of gravity, and is almost identical to GR.\\
	\section{The Equivalence and the Effacing principle in General Relativity}
	\label{s:EqEfGR}
	\subsection{A coordinate system adapted for the internal problem}
	\label{ss:coordsys}
	An asymptotic series of solutions based on Newtonian initial data can be found as a function of multipole moments of classical matter using a formalism developed by Schutz and Futamase \cite{Schutz1983}, by using two co-moving and scaled 'body zone' coordinate systems for each of the constituents of the binary, compared to an observer at Minkowskian asymptotic infinity.\\
	
	The formalism involves defining a scaled time coordinate $ s $ (different from $ \tau $ as was used in \cite{Futamase1987}), with respect to the time coordinate $ t $ used by an asymptotic observer at rest, with respect to the center of mass of the binary system, 
	\begin{eqnarray}
	s &=& \epsilon\,t, \label{newtym}
	\end{eqnarray}
	which is well suited for the internal problem of the two bodies, given they are non-spinning or slowly spinning, as was found in \cite{Schutz1983}. $ \epsilon $ is the post Newtonian parameter given by $ \frac{\left|v_{orb}\right|}{c} $, with $ \left|v_{orb}\right| $ being the absolute value of the orbital velocity, and $ c $ being the speed of light (unity in the current study). The re-parameterized time $ s $ is also taken as the proper time of a co-moving observer along either sources. \\
	
	A transformation into the 'body zone coordinate system' involves first setting up of a Fermi-Walker coordinate system along the center of mass of two compact objects \cite{Futamase1987,Itoh2000,Futamase2007}, whose centers of masses follow two world lines, denoted by 3-vectors $ z_L^i\left(s\right) $, for $ L\,=\,1,\,2 $. Under such a transformation, various components of a contravariant tensor, as defined initially in the asymptotic observer's frame (denoted in primed lowercase Latin), transform into the co-moving frame as follows
	\begin{eqnarray}
	T_L^{s\,s} &=& T_L^{s\,s} \label{emfmss}\\
	T_L^{s\,i} &=& \epsilon^2 T_L^{s\,i'} + v_L^{i} T_L^{s\,s} \label{emfmsi} \\
	T_L^{i\,j} &=& \epsilon^4 T_L^{i'\,j'} + 2\epsilon^2 v_L^{\left(i\right.} T_L^{\left.j'\right) s} + v_L^i \,v_L^j T_L^{s\,s} \label{emfmij}
	\end{eqnarray}
	The 'smallness' parameter $ \epsilon $ is the ratio of the average orbital velocities and the speed of light (unity in the current study), as agreed upon by an observer at asymptotic infinity, who is static with respect to the center of mass of the two bodies. Then the 3-velocity of the body $ v_L^i\left(\,=\,\frac{dz^i_L}{ds}\right) $, as measured by an external asymptotic observer, is of order unity.\\
	
	The next step involves defining a specially scaled spatial coordinate system for the internal problem. For compact objects whose mass-radii ratios approach unity, given the orbital separation is held fixed, both mass and radius scale as $ \epsilon^2 $, if one were to utilize a Newtonian/Keplerian initial data. Consequently, mass-energy densities scale as $ \epsilon^{-4} $. Hence, the internal problem can be solved in coordinate systems $ X_L^i $ (as defined in \cite{Futamase1987,Itoh2000,Futamase2007}) that is scaled by the parameter $ \epsilon^2 $, such that
	\begin{eqnarray}
	X_L^{i} &\equiv& \frac{x^{i}-z_L^{i}\left(s\right)}{\epsilon^2}. \label{scalcoord}
	\end{eqnarray}
	Under the above transformation, various contravariant components of the energy-momentum tensor $ T^{\mu\nu} $ scale as follows in the body zone
	\begin{eqnarray}
	T_L^{s\,s} &=& \mathcal{O}\left(\epsilon^{-2}\right) \label{emscalss} \\
	T_L^{s\,i'} &=& \mathcal{O}\left(\epsilon^{-4}\right) \label{emscalsi}\\
	T_L^{i'\,j'} &=& \mathcal{O}\left(\epsilon^{-8}\right) \label{emscalij}
	\end{eqnarray}
	which completes the set of transformations necessary to solve the internal problem. It is to be noted that, as seen in the special coordinate systems (the body zone coordinates), the radii of either compact objects are unity. 
	\subsection{Equations of motion}
	\label{ss:GReom}
	Using the 'special' coordinate system defined in the previous section, a sequence of solutions can be found for the internal problem. The relaxed form of the field equations of GR, with a matter source tensor $ T^{\mu\nu}_m $, in the asymptotically Minkowskian observer's coordinate system are given as follows
	\begin{eqnarray}
	\partial_{\alpha\,\beta}H^{\mu\alpha\nu\beta} &=& 16\,\pi\, \left(-g\right) \left(T^{\mu\nu}_m + t_{LL}^{\mu\nu}\right) \label{llgr} \\
	H^{\mu\alpha\nu\beta} &=& \mathfrak{g}^{\alpha\beta}\mathfrak{g}^{\mu\nu} - \mathfrak{g}^{\alpha\nu}\mathfrak{g}^{\beta\mu} \label{maxRie}
	\end{eqnarray}
	$ \mathfrak{g}^{\mu\nu} $ is the square root of the determinant weighed contravariant metric $ \left(\sqrt{-g}\,g^{\mu\nu}\right) $. In this picture of GR, the dynamical variable is not the metric $ g^{\mu\nu} $, but the contravariant metric density $ \mathfrak{g}^{\mu\nu} $ (also known as the gothic metric in PN literature), which propagates on a Minkowski background $ \eta^{\mu\nu} $. An apparent separation of the metric of a maximally symmetric space-time (corresponding to the tensor $ H^{\mu\alpha\nu\beta} $), from the part of the metric that leads to curvature of space-time (the effective energy-momentum pseudo-tensor of gravitation $ t^{\mu\nu}_{LL} $), is possible in this formalism. The exact form is given by
	\begin{eqnarray}\label{EMSDLandauLifshitz}
	&&16\,\pi\,\left(-g\right)\,t_{\mathrm{LL}}^{\alpha \beta} = \partial_{\lambda} \mathfrak{g}^{\alpha \beta} \partial_{\mu} \mathfrak{g}^{\lambda \mu}-\partial_{\lambda} \mathfrak{g}^{\alpha \lambda} \partial_{\mu} \mathfrak{g}^{\beta \mu}\nonumber \,+\,\frac{1}{2} \mathfrak{g}^{\alpha \beta} \mathfrak{g}_{\lambda \mu} \partial_{\rho} \mathfrak{g}^{\lambda \nu} \partial_{\nu} \mathfrak{g}^{\mu \rho}
	-\mathfrak{g}^{\alpha \lambda} \mathfrak{g}_{\mu \nu} \partial_{\rho} \mathfrak{g}^{\beta \nu} \partial_{\lambda} \mathfrak{g}^{\mu \rho}\nonumber\\
	&&\,-\,\mathfrak{g}^{\beta \lambda} \mathfrak{g}_{\mu \nu} \partial_{\rho} \mathfrak{g}^{\alpha \nu} \partial_{\lambda} \mathfrak{g}^{\mu \rho}\,+\,\mathfrak{g}_{\lambda \mu} \mathfrak{g}^{\nu \rho} \partial_{\nu} \mathfrak{g}^{\alpha \lambda} \partial_{\rho} \mathfrak{g}^{\beta \mu} \,+\, \frac{1}{8}\left(2 \mathfrak{g}^{\alpha \lambda} \mathfrak{g}^{\beta \mu}\,-\,\mathfrak{g}^{\alpha \beta} \mathfrak{g}^{\lambda \mu}\right)\left(2 \mathfrak{g}_{\nu \rho} \mathfrak{g}_{\sigma \tau} \,-\, \mathfrak{g}_{\rho \sigma} \mathfrak{g}_{\nu \tau}\right) \partial_{\lambda} \mathfrak{g}^{\nu \tau} \partial_{\mu} \mathfrak{g}^{\rho \sigma} \nonumber\\
	\\
	&&\mathfrak{g}_{\mu\nu} \equiv \frac{g_{\mu\nu}}{\sqrt{-g}}
	\end{eqnarray}
	which has at most one derivative of $ \mathfrak{g}^{\alpha\beta} $. The purpose of the metric $ g $ in this formalism is solely to raise and lower indices of the auxiliary variable $ \mathfrak{g}^{\alpha\beta} $. It is to be noted that the above definition is neither covariant nor unique. A covariant proposition for the Landau-Lifshitz energy-momentum pseudo-tensor can be found in \cite{Harte:2014}.\\
	
	From the old days of Schwarzschild \cite{Schwarzschild:1916uq}, the relaxed form of the field equations, or the Landau-Lifshitz formalism, helps one to find exact solutions of the highly non-linear field equations of GR, as was also explained clearly in \cite{Fromholz:2013hka}.\\
	
	With regards to observable effects, the biggest accomplishment of the relaxed formalism has been to obtain a series of potentials known as the Post-Newtonian expansion of GR in literature, as in \cite{Einstein1949OnTM,Infeld1949}, and later systematized in \cite{Chandra1965,*Chandra1967,*Chandra1969I,*Chandra1969II,*Chandra1970}. It involves defining tensor potentials $ \mathfrak{h}^{\mu\nu} $, given by
	\begin{eqnarray}\label{redpot}
	\mathfrak{h}^{\mu\nu} &=& \eta^{\mu\nu}-\mathfrak{g}^{\mu\nu}
	\end{eqnarray}
	along with a constraint
	\begin{eqnarray}\label{harmg}
	\partial_{\mu} \mathfrak{h}^{\mu\nu} &=& 0,
	\end{eqnarray}
	which restricts the possible class of coordinates (or gauges), in which one attempts to find solutions for $ \mathfrak{h}^{\mu\nu} $. In PN and GW literature, the constraint is known as the harmonic gauge, in which the coordinate chart $ x^\mu $ follow four massless and homogeneous wave equations
	\begin{eqnarray}
	\Box\,x^\mu &=& 0 \label{harmchart}
	\end{eqnarray}
	where $ \Box $ is the d'Alembert operator or $ \eta^{\mu\nu}\partial_{\mu}\partial_{\nu} $. Such a class of coordinates are consequently known as harmonic coordinate systems, and the dynamical equations for $ \mathfrak{h}^{\mu\nu} $ take the following form
	\begin{eqnarray}
	\Box \mathfrak{h}^{\mu\nu} &=& -16\,\pi\,\Lambda^{\mu\nu} \label{LLh} \\
	\Lambda^{\mu\nu}&=& \Theta^{\mu\nu} + \partial_{\alpha\,\beta}\chi^{\mu\nu\alpha\beta} \label{effemp} \\
	\Theta^{\mu\nu} &=& \left(-g\right) \left(T^{\mu\nu} \,+\, t_{LL}^{\mu\nu} \,+\, t_H^{\mu\nu}\right)\label{effem}\\
	\chi^{\mu\nu\alpha\beta} &=& \frac{1}{16\,\pi} \left(\mathfrak{h}^{\alpha\nu} \mathfrak{h}^{\beta\mu} - \mathfrak{h}^{\alpha\beta}\mathfrak{h}^{\mu\nu}\right) \label{surfterm}
	\end{eqnarray}
	Eq. (\ref{LLh}) can be solved for $ \mathfrak{h}^{\mu\nu} $ in a series of multipole moments of the energy-momentum pseudo-tensor $ \Lambda^{\mu\nu} $. It is to be noted that the RHS of Eq. (\ref{surfterm}) manifests as a total derivative term in the field equations and can be transformed into boundary terms using Gauss's law. Choice of the boundary condition fixes the form of the Green's function. For the case in study, the no incoming radiation from past null Minkowskian infinity boundary condition will be used, which was given in Sec. 4.1 of \cite{Futamase2007} as follows
	\begin{eqnarray}\label{boundcon}
	\lim\limits_{s\,\rightarrow\,r\,,\,r\,\rightarrow\,\infty}\,\,\left[\frac{\partial}{\partial\,r}\left(r\,\mathfrak{h}^{\mu\nu}\right)\,+\,\frac{\partial}{\partial\,s}\left(r\,\mathfrak{h}^{\mu\nu}\right)\right] &=& 0
	\end{eqnarray}
	where $ r $ is the radial coordinate distance from the COM of a compact source to a field point. Enforcing the condition (\ref{boundcon}), the effect of the terms in the RHS of Eq. (\ref{surfterm}) on the solution sequence $ \mathfrak{h}^{\mu\nu} $ vanishes. Various components of the tensor potentials $ \mathfrak{h}^{\mu\nu} $ till $ \mathcal{O}\left(\epsilon^6\right) $, in the body zone coordinate system, are given as follows \cite{Futamase1987,Itoh2000,Futamase2007}
	\begin{eqnarray}
	\mathfrak{h}_\mathcal{B}^{s\,s} &=& 4\, \epsilon^4\, \sum_{L=1,\,2} \, \left(\frac{P^s_L}{\left|Z^i_L\right|} + \epsilon^2\,\frac{D^k_L\,Z^k_L}{\left|Z^i_L\right|^3} \right) + \mathcal{O}\left(\epsilon^{8}\right) \label{fhssmul} \\
	\mathfrak{h}_\mathcal{B}^{s\,i} &=& 4 \, \epsilon^4 \sum_{L=1,\,2} \, \left(\frac{P_L^i}{\left|Z^i_L\right|} + \epsilon^2\,\frac{J_L^{k\,i} Z_L^k}{\left|Z^i_L\right|^3}\right) + \mathcal{O}\left(\epsilon^8\right) \label{fhsimul} \\
	\mathfrak{h}_\mathcal{B}^{i\,j} &=& 4 \, \epsilon^2 \, \sum_{L=1,\,2} \left(\frac{Y_L^{i\,j}}{\left|Z^i_L\right|} + \epsilon^2 \, \frac{Y_L^{k\,i\,j} \, Z_L^k}{\left|Z^i_L\right|^3}\right) + \mathcal{O}\left(\epsilon^6\right) \label{fhijmul}
	\end{eqnarray}
	where $ Z^i_L\left(s\right)\,=\,x^i\,-\,z^i_L\left(s\right) $ is the displacement 3-vector of the field point to the COM of the source. In the current formalism, the above components of the tensor potential are sufficient to obtain equations of motion of the two sources till the first PN order. The monopole and dipole moments, respectively, of various components of $ \Lambda^{\mu\nu} $ are defined as follows
	\begin{eqnarray}
	P_L^{s} &=& \lim\limits_{\epsilon\,\rightarrow\,0}\, \int_{\mathcal{B}_L} d^3X_L \, \Lambda^{s\,s} \label{monoss}\\
	P^i_L &=& \lim\limits_{\epsilon\,\rightarrow\,0}\, \int_{\mathcal{B}_L} d^3X_L \, \Lambda^{s\,i} \label{monosi} \\
	Y^{i\,j}_L &=& \lim\limits_{\epsilon\,\rightarrow\,0}\, \int_{\mathcal{B}_L} d^3X_L \, \Lambda^{i\,j} \label{monoij}\\
	D^i_L &=& \lim\limits_{\epsilon\,\rightarrow\,0}\, \int_{\mathcal{B}_L} d^3X_L \, \Lambda^{s\,s} \, X^i_L \label{diss}\\
	J^{i\,j}_L &=& \lim\limits_{\epsilon\,\rightarrow\,0}\, \int_{\mathcal{B}_L} d^3X_L \, \Lambda^{s\,i} X^j_L \label{disi} \\
	Y^{i\,j\,k}_L &=& \lim\limits_{\epsilon\,\rightarrow\,0}\, \int_{\mathcal{B}_L} d^3X_L \, \Lambda^{i\,j} \, X^k_L \label{diij}
	\end{eqnarray}
	where the various pre-factors of $ \epsilon^n $ for $ n\,=\,2,\,4,\,8 $ for the various multipole moment definitions have been removed, compared to \cite{Futamase1987,Itoh2000,Futamase2007}, in order to avoid any confusion regarding the scaling of various components of a rank two tensor $ \Lambda^{s\,s} $ under the transformations (\ref{newtym}) and (\ref{scalcoord}).\\
	
	The equations of motion at various PN orders come from the definition of a quasi-local four-momenta of the COM of each of the bodies in the binary as defined in \cite{Futamase1987,Itoh2000,Futamase2007,Giulini:2018} as
	\begin{eqnarray}\label{fmom}
	P^\mu_L\left(s\right) &=& \epsilon^2\,\int_{\mathcal{B}_L}\,d^3X_L\,\Lambda^{s\,\mu},
	\end{eqnarray}
	and the conservation law
	\begin{eqnarray}
	\Lambda^{\mu\nu}_{\quad\,,\,\nu} &=& 0,
	\end{eqnarray}
	from which one obtains the evolution equation of the four-momenta $ P^\mu_L $ as surface integrals of the energy-momentum pseudo-tensor, over the boundary of the body zone of the $ L^{th} $ object, as was obtained in \cite{Futamase1987,Itoh2000,Futamase2007}:
	\begin{eqnarray}
	\frac{dP^\mu_L}{ds} &=& -\epsilon^{-4}\,\oint_{\partial\mathcal{B}_L}\,dS_k\,\Lambda^{k\,\mu} + \epsilon^{-4}\,v^k_L\,\oint_{\partial\mathcal{B}_L}\,dS_k\,\Lambda^{s\,\mu}, \label{momevol}\nonumber\\
	\end{eqnarray}
	where $ dS_k $ is an infinitesimal unit vector normal to the 2-sphere $ \partial\mathcal{B}_L $ that is the boundary of the body zone $ \mathcal{B}_L $. The 3-momentum vs 3-velocity relationship is given by
	\begin{eqnarray}
	P^i_L &=& P^s_L\,v^i_L + Q^i_L + \mathcal{O}\left(\epsilon^2\right) \\
	Q^i_L &=& \epsilon^{-4} \oint_{\partial\mathcal{B}_L} dS_k \left(\Lambda^{s\,k} - v_L^k\, \Lambda^{s\,s}\right) X_L^i \label{qi}
	\end{eqnarray}
	Newtonian like equations of motion involve evolution equations of 3-velocities, which are the 3-accelerations. They were given in \cite{Futamase1987,Itoh2000,Futamase2007} as
	\begin{eqnarray}\label{vevol}
	P^s_L \, \frac{dv^i_L}{ds} &=& -\epsilon^{-4} \oint_{\partial\mathcal{B}_L} dS_k\, \Lambda^{k\,i} + \epsilon^{-4} \, v^k_L \oint_{\partial\mathcal{B}_L}dS_k \,\Lambda^{s\,i} \nonumber\\
	&&+ \epsilon^{-4}\,v^i_L \left(\oint_{\partial\mathcal{B}_L}dS_k\, \Lambda^{k\,s} - v_L^k \oint_{\partial\mathcal{B}_L}dS_k \,\Lambda^{s\,s}\right)\nonumber\\
	&& - \frac{dQ^i_L}{ds},
	\end{eqnarray}
	If one chooses a definition of the mass $ M_L $ of the $ L^{th} $ body as
	\begin{eqnarray}
	M_L &=& \lim\limits_{\epsilon\,\rightarrow\,0}\,P^s_L,\label{grmassdef}
	\end{eqnarray}
	then the equations of motion of the first body, for example, till first PN order were obtained in \cite{Futamase1987,Itoh2000,Futamase2007} as
	\begin{eqnarray}\label{GReom}
	M_1\,\frac{dv_1^i}{ds} &=& -\frac{M_1\,M_2}{r_{1\,2}^{\,2}}\,n^i\,+\,\epsilon^2\,\frac{M_1\,M_2}{r_{1\,2}^2}\,\left[\left(-v_1^2-2\,v_2^2\right.\right.\nonumber\\
	&&\left.\left.+\frac{3}{2}\,\left(\hat{n}\,\cdot\textbf{v}_2\right)^2+4\,\left(\textbf{v}_1\,\cdot\,\textbf{v}_2\right)+\frac{5\,M_1}{r_{1\,2}}+\frac{4\,M_2}{r_{1\,2}}\right)\,n^i\right.\nonumber\\
	&&\left.\,+\,\left\{4\,\left(\hat{n}\,\cdot\,\textbf{v}_1\right)\,-3\,\left(\hat{n}\,\cdot\,\textbf{v}_2\right)\right\}\,\left(v_1^i-v_2^i\right)\right]\,+\,\mathcal{O}\left(\epsilon^4\right)\nonumber\\
	\end{eqnarray}
	where $ n^i\,\left(\,\equiv\,\hat{n}\right) $ is a unit vector pointing from the COM of the first body zone to the COM of the second body zone, and $ r_{1\,2} $ is the distance between the COM of the two body zones or the orbital separation. Mass multipole moments of $ \Lambda^{\mu\nu} $, as seen by an observer at rest (w.r.t. the COM of the binary system) at asymptotic infinity, go as $ \left(\text{mass}\right)\,\times\,\left(\text{average radius of mass distribution}\right)^\ell $. Using Newtonian initial data, mass multipole objects appearing in the equations of motion of the binary (\ref{vevol}) scale as $ \mathcal{O}\left(\epsilon^{2\,\ell\,+\,2}\right) $. In terms of the classical notion of force on a particle,
	\begin{eqnarray}\label{GReffforce}
	F^i &=& F^i_{N} \,+\,\epsilon^2\,F^i_{1PN} \,+\, \mathcal{O}\left(\epsilon^4\right)
	\end{eqnarray}
	corresponding to the first and second term of the RHS of Eq. (\ref{GReom}), respectively. If one simply uses the first term of (\ref{GReom}), as a crude approximation for the trajectory of two slow spinning compact objects in the early inspiral phase, one obtains the Newtonian force on the first body. The dynamics of the first body then is independent of its own mass and depends only on the Newtonian potential generated by the second body, which is a consequence of the equivalence principle. The generalization of the consequences of the equivalence principle till first PN order comes through the effacing principle, where the internal/structural details of either body, and their effect on the external gravitational field, do not factor into their trajectories around/towards each other. The compact objects still behave like point particles moving along the geodesics of an external gravitational field. Self force effects through the appearance of velocity dependent terms appear at the first PN order. Although from the first PN order onward the dynamics of the first body zone is dependent on the mass $ M_1 $ enclosed in it, as was defined in Eq. (\ref{grmassdef}), there exist no multipole objects other than the monopole moments of $ \Lambda^{\mu\nu} $, essentially making the two compact objects behave like massive but point particles (not to be confused with test masses).\\
	
	The effects of spin in the equations of motion are dependent on the scaling of the current multipole moments of $ \Lambda^{\mu\nu} $. In the slow rotation approximation, the internal velocities scale as $ \mathcal{O}\left(\epsilon\right) $. Hence current multipole moments go as $ \left(\text{mass}\right)\,\times\,\left(\text{average radius of mass distribution}\right)^\ell\,\times\,\left(\text{velocity of internal motion}\right) $, which is $ \mathcal{O}\left(\epsilon^{2\,\ell\,+\,3}\right) $, implying that the spin-orbit coupling force is of the form $ \left(\text{mass}\right)\,\times\,\left(\text{orbital velocity}\right)\,\times\,\left(\text{spin}\right) $, or $ \mathcal{O}\left(\epsilon^{2\,\ell\,+\,4}\right) $. It is also to be noted that the scalings change when the velocity of internal motion cannot be ignored for rapidly rotating constituents of the binary. The time scaling in Eq. (\ref{newtym}), then changes to $ s\,\equiv\,\epsilon^{-2}\,t $, as was described in \cite{Futamase1987}.\\
	
	The above described consequence of GR is a way one may choose to interpret the effects of the equivalence assumption. But this kind of a consequence is not unique to GR, as will be shown in the following section.
	\section{First PN equations of motion in $ f(R) $ gravity and the Effacing principle}
	\label{s:EqEffR}
	\subsection{Relaxing the $ f\left(R\right) $ field equations}
	\label{relfR}
	The field equations of $ f\left(R\right) $ theories of gravity can be found by varying the action in Eq. (\ref{fRaction}). It is however, for the sake of understanding the physical consequences of such theories, better to frame the field equations in an Einsteinian way; such that possible observable deviations from GR can be expressed as an effective energy-momentum tensor, that is different from the ordinary matter energy-momentum tensor. In the latter way, the contravariant field equations are expressed as \cite{Capozziello2010}
	\begin{eqnarray}
	&&G^{\mu\nu} = \frac{8\,\pi}{f'}\,\left(T_m^{\mu\nu}\,+\,T^{\mu\nu}_{eff}\right) \label{fRfield}\\
	&&3\,g^{\mu\nu}\,\nabla_{\mu}\,\nabla_{\nu}\,f'\,+\,f'\,R\,-\,2\,f = 8\,\pi\,T_m \label{scalfield} \\
	T^{\mu\nu}_{eff} &\equiv& \nabla^\mu\,\nabla^\nu\,f'\,+\,\frac{g^{\mu\nu}}{2}\,\left(f\,-\,R\,f'\right)\,-\,g^{\mu\nu}\,g^{\alpha\beta}\,\nabla_{\alpha}\,\nabla_{\beta}\,f' \nonumber\\
	\\
	f'\left(R\right) &\equiv& f' \,=\, \frac{df\left(R\right)}{dR} \qquad\,T_m\,=\,g_{\alpha\beta}\,T_m^{\alpha\beta}\qquad \nabla^{\mu} \equiv g^{\mu\nu}\,\nabla_\nu\nonumber\\
	\end{eqnarray}
	where the subscript $ m $ has been utilized to distinguish the classical matter energy-momentum tensor from the effective one. $ G^{\mu\nu} $ is the Einstein tensor of GR, and $ \nabla_{\mu} $ is the covariant derivative associated with the metric $ g^{\mu\nu} $.\\
	
	The above system of partial differential equations are extremely non-linear in the metric and the scalar field $ f' $, and needs some form of relaxing before a solution, similar to the PN expansion in GR, can be generated as functions of multipole moments of an effective source given by the RHS of Eq. (\ref{fRfield}).\\
	
	There has been a number of analysis on $ f\left(R\right) $ theories of gravity. In the cosmological context and local gravity constraints, a review can be found in \cite{DeFelice2010,Nojiri:2010wj,Nojiri:2017ncd}. Early solutions for spherically symmetric distributions of fluids can be found in \cite{Schmidt1986,Teyssandier1990,Cognola:2005de,Olmo2005,Berry2011}. An analysis using null tetrads to obtain conditions for deviations to a curvature invariant object (Weyl scalar) $ \Psi_2 $ was done in \cite{ALVES2009401}. Perturbations about a completely homogeneous and isotropic space-time was performed in \cite{Bean2007}, whereas perturbation studies about a spherically symmetric space-time using the standard metric PN expansion (without using the gothic metric density) was done in \cite{Capozziello2007}. Stability of solutions about perturbations in general space-times for various deviation parameters was shown in \cite{Bazeia2007}.\\
	
	In the spirit of the Landau-Lifshitz formalism, an auxiliary metric $ \mathfrak{\tilde{g}}^{\mu\nu} $, similar to the gothic metric $ \mathfrak{g}^{\mu\nu} $ of GR, but different from \cite{Bean2007,Capozziello2007} can be defined as
	\begin{eqnarray}\label{newgoth}
	\mathfrak{\tilde{g}}^{\mu\nu} &=& f'\left(R\right)\,\sqrt{-g}\,g^{\mu\nu}
	\end{eqnarray}
	which will be the dynamical variable in the current study, propagating in a Minkowski background $ \eta^{\mu\nu} $. It is to be noted that, under linearization of the field equations of $ f\left(R\right) $ theories of gravity with a well defined polynomial expansion about $ R\,=\,0 $, the RHS of Eq. (\ref{newgoth}) reduces to the transverse-traceless metric deviation tensor $ \bar{h}^{\mu\nu} $ in \cite{Berry2011}.\\
	
	Using the new definition, the field equations (\ref{fRfield}) can be written as follows
	\begin{eqnarray}
	\partial_{\alpha\,\beta} H^{\alpha\mu\beta\nu}  &=& 16\,\pi\,\left(-g\right)\,f'\left(R\right) \,\left(T_m^{\mu\nu} \,+\, t_{eff}^{\mu\nu}\,+\, t^{\mu\nu}_{LL}\right),\label{fRLL}\\
	16\,\pi\,\left(-g\right)\,f'\,t^{\mu\nu}_{eff} &=& \sqrt{-g}\,\left(f\,-\,R\,f'\right)\,\tilde{\mathfrak{g}}^{\mu\nu} \,+\, \frac{3}{\left(f'\right)^2}\,\left(\tilde{\mathfrak{g}}^{\mu\alpha}\,\tilde{\mathfrak{g}}^{\nu\beta}\,-\,\frac{1}{2}\,\tilde{\mathfrak{g}}^{\mu\nu}\,\tilde{\mathfrak{g}}^{\alpha\beta}\right)\,\partial_{\alpha}f'\,\partial_{\beta}f' \label{teffgenfR}\\
	f &\equiv& f\left(R\right)\,;\, \qquad\qquad f' \,\equiv\,f'\left(R\right)
	\end{eqnarray}
	where the forms of $ H^{\alpha\mu\beta\nu} $ and $ t^{\mu\nu}_{LL} $ remain the same as in GR (with $ \mathfrak{g} $ replaced by $ \tilde{\mathfrak{g}} $) as was defined in Eqs. (\ref{maxRie}) and (\ref{EMSDLandauLifshitz}), whereas $ t^{\mu\nu}_{eff} $ is seen to be a quadratic function of only first derivatives of $ f'\left(R\right) $. The functional forms of $ f\left(R\right) $ and $ f'\left(R\right) $ can be written as a power series about $ R\,=\,0 $ as
	\begin{eqnarray}
	f\left(R\right) &=& R\,+\,\frac{f''\left(0\right)}{2}\,R^2 \label{fRform}\\
	f'\left(R\right) &=& 1\,+\,f''\left(0\right)\,R \label{fpRform}
	\end{eqnarray}\\
	where the coefficients of expansion $ f'\left(0\right) $ is taken to be unity to recover GR at the $ R\,=\,0 $ limit, and $ f''\left(0\right) $ will be taken to be negative for the course of this article, following \cite{Berry2011,Schmidt1986,Teyssandier1990,Olmo2005}. The negative choice for the constant $ f''\left(0\right) $ also frames Eq. (\ref{scalfield}), after making the choices (\ref{fRform}) and (\ref{fpRform}), into a Klein-Gordon equation that is satisfied by the massive scalar field.\\
	
	Perturbing the configuration $ \left(\mathfrak{\tilde{g}},\,R\right) $ about a Minkowski space-time $ \eta^{\mu\nu} $, Eq. (\ref{fRLL}) can now be perturbatively solved as functions of multipole moments of net energy-momentum pseudo-tensors $ \left(\tilde{\Lambda}^{\mu\nu},\,\tilde{\Lambda}\right) $. One can define  a tensor potential $ \tilde{\mathfrak{h}}^{\mu\nu} $, similar to $ \mathfrak{h}^{\mu\nu} $ of GR, that propagates on a Minkowski space-time $ \eta^{\mu\nu} $
	\begin{eqnarray}
	\mathfrak{\tilde{h}}^{\mu\nu} &=& \eta^{\mu\nu}\,-\,\mathfrak{\tilde{g}}^{\mu\nu} ,\label{tenspert}
	\end{eqnarray}
	and use a conformal-Lorenz gauge condition (referred to as the "Lorentz" gauge condition in \cite{Mirshekari:2013vb}, and as conformal-harmonic condition in \cite{Kopeikin:2020rxp})
	\begin{eqnarray}\label{conlor}
	\mathfrak{\tilde{h}}^{\mu\nu}_{\quad\,,\,\nu} &=& 0.
	\end{eqnarray}
	Since there is the Ricci scalar manifesting as a scalar field, one needs to define a dynamical variable that propagates on a Ricci flat background, and has a "proper" scaling relationship with the trace of the classical energy-momentum tensor (which may comprise of a pair of slowly spinning fluids, gravitational mass monopoles, or a combination of both). Noticing that $ \Theta^{\mu\nu} $ has an overall factor of $ \left(-g\right) $ in the RHS of Eq. (\ref{effem}), such that $ \Box\,\mathfrak{h}^{\mu\nu} $ is related to the classical energy-momentum tensor $ T_m^{\mu\nu} $ with an overall $ \left(-g\right) $ factor. One may choose to structure the trace of the field equations (\ref{scalfield}) in a manner that connects the d'Alembert operated scalar density, say $ \Box\,\mathfrak{R} $, to the trace of $ T_m^{\mu\nu} $ with an overall $ \left(-g\right) $ factor.\\
	
	In the action, the square root of the negative of metric determinant multiplied with the Ricci scalar is akin to the energy-density (if one uses geometrized units). Hence the scalar variable, like the tensor variable $ \mathfrak{g}^{\mu\nu} $ in GR and $ \tilde{\mathfrak{g}}^{\mu\nu} $ in $ f\left(R\right) $, is changed to be the following
	\begin{eqnarray}
	\mathfrak{R} &=& \sqrt{-g}\,R\label{scalpert}
	\end{eqnarray}
	One obtains the relaxed dynamics of $ \mathfrak{\tilde{h}}^{\mu\nu} $ as
	\begin{eqnarray}\label{hfrakdyn}
	\Box \mathfrak{\tilde{h}}^{\mu\nu} &=& 16\,\pi\,\tilde{\Lambda}^{\mu\nu}\\
	\tilde{\Lambda}^{\mu\nu} &=& \left(-g\right)\,\left[T^{\mu\nu}_m\,\left(1\,-\,\frac{f''\,\,\mathfrak{R}}{\sqrt{-g}}\right)\,+\,t_{LL}^{\mu\nu}\,+\,t_H^{\mu\nu}\,+\,t_{eff}^{\mu\nu}\right].\label{EMSDnetfR}
	\end{eqnarray}
	with the following conservation law being satisfied by $ \tilde{\Lambda}^{\mu\nu} $
	\begin{eqnarray}
	\tilde{\Lambda}^{\mu\nu}_{\,\,\,,\,\mu} &=& 0\,\,\,. \label{effemcons}
	\end{eqnarray}
	$ t_{eff}^{\mu\nu} $ is comprised of various products of $ \tilde{\mathfrak{h}}^{\mu\nu} $, $ \mathfrak{R} $, and first derivatives of $ \mathfrak{\tilde{h}}^{\mu\nu} $ and $ \mathfrak{R} $; whose truncated form till the quadratic order of $ f''\left(0\right) $ is given as follows
	\begin{eqnarray}
	&&16\,\pi\,\left(-g\right)\,t^{\mu\nu}_{eff} = \,\frac{f''\,\,\mathfrak{R}^2}{\sqrt{-g}}\,\,\eta^{\mu\nu}\,+\,\frac{3\,\left(f''\right)^2}{4\,\left(-g\right)}\,\left[2\,\left(\eta^{\mu\alpha}\,\eta^{\nu\beta}\,+\,\eta^{\mu\beta}\,\eta^{\nu\alpha}\right.\right.\nonumber\\
	&&\left.\left.\,-\,\eta^{\mu\nu}\,\eta^{\alpha\beta}\right)\,\mathfrak{R}\,\partial_{\alpha}\mathfrak{R}\,\partial_\beta\tilde{\mathfrak{h}}\,+\,\left(\eta^{\mu\alpha}\,\eta^{\nu\beta}\,-\,\frac{1}{2}\,\eta^{\mu\nu}\,\eta^{\alpha\beta}\right)\,\left(4\,\partial_{\alpha}\mathfrak{R}\,\partial_\beta\mathfrak{R}\,+\,\mathfrak{R}^2\,\partial_{\alpha}\tilde{\mathfrak{h}}\,\partial_{\beta}\tilde{\mathfrak{h}}\right)\right]\nonumber\\
	&&\,+\,\mathcal{O}\left[\left(\frac{f''}{\sqrt{-g}}\right)^3\right]\label{teffrel}\\
	&&\tilde{\mathfrak{h}}\,\equiv\,\eta_{\mu\nu}\,\tilde{\mathfrak{h}}^{\mu\nu}
	\end{eqnarray}
	It is to be noted that under the transformation (\ref{newtym}), the covariant Minkowski metric, using the $ \left(+\,,\,-\,,\,-\,,\,-\right) $ signature becomes
	\begin{eqnarray}\label{scaledminkowski}
	\eta_{\mu\nu} &\equiv& diag\left(\epsilon^{-2},\,-1,\,-1,\,-1\right)
	\end{eqnarray}
	therefore $ \left|\det\left(\eta_{\mu\nu}\right)\right|\,\equiv\,\left(-g\right) $ scale as $ \epsilon^{-2} $. It is to be noted that the trace of the new (and the old) gothic metric density deviation $ \tilde{\mathfrak{h}} $ scale as $ \epsilon^2 $ under the transformation (\ref{newtym}), compared to the individual components of $ \tilde{\mathfrak{h}}^{\mu\nu} $ at the spinless or slowly spinning limit, which scale as $ \epsilon^4 $. It is important to note the coefficients of $ \left(-g\right)^{-1} $ in the second term of Eq. (\ref{teffrel}) (inside the square brackets) as a whole are sub-leading (in powers of $ \epsilon $ in the PN expansion), compared to the first term of Eq. (\ref{teffrel}) which is in turn sub-leading with respect to $ T_m^{\mu\nu} $. Notice that the first appearance of a non GR  term in $ \tilde{\Lambda}^{\mu\nu} $ in the $ \epsilon $ sequence is a non-minimal coupling between the Ricci scalar and the classical matter energy-momentum tensor along with a term quadratic in $ \mathfrak{R} $.\\
	
	$ t_H^{\mu\nu} $ appears in the PN formulation of GR as well, whose form was given in Eqs. (\ref{effemp}) and (\ref{surfterm}) through the term $ \chi^{\mu\nu\alpha\beta}_{\quad\,,\,\alpha\beta} $, with $ \mathfrak{h}^{\mu\nu} $ replaced by $ \tilde{\mathfrak{h}}^{\mu\nu} $. The total derivative term allows one to put $ t^{\mu\nu}_H $ to zero at the boundaries of integration because of the specific boundary condition choice of no incoming tensor radiation from past null Minkowskian infinity.\\
	
	Under the transformations in Eqs. (\ref{newgoth}) and (\ref{scalpert}), the trace of the field equations, that is Eq. (\ref{scalfield}), reduces to the following
	\begin{eqnarray}\label{newtracepow}
	&&\Box\,\mathfrak{R}\,+\,\sqrt{-g}\,\gamma^2\,\mathfrak{R} = -8\,\pi\,\left(-g\right)\,\gamma^2\,\tilde{\Lambda}\\
	&&\tilde{\Lambda} = \,T_m\,+\,f''\left(0\right)\,\left\{\frac{T_m\,\mathfrak{R}}{3\,\sqrt{-g}}+\,\frac{1}{8\,\pi\,\left(-g\right)}\left(\frac{\mathfrak{R}^2}{3}\,-\,\partial_\mu\mathfrak{R}\,\partial^\mu\tilde{\mathfrak{h}}\,-\,\frac{1}{2}\,\mathfrak{R}\,\partial_{\mu}\tilde{\mathfrak{h}}^{\alpha\beta}\,\partial^{\mu}\tilde{\mathfrak{h}}_{\alpha\beta}\right.\right.\nonumber\\
	&&\left.\left.\,-\,\frac{1}{4}\,\mathfrak{R}\,\partial_{\mu}\tilde{\mathfrak{h}}\,\partial^\mu\tilde{\mathfrak{h}}\,-\,\frac{1}{2}\,\mathfrak{R}\,\Box\tilde{\mathfrak{h}}\right)\right\}\,+\,\mathcal{O}\left(\left[f''\left(0\right)\right]^2\right)\label{effemtilde}\\
	&&\gamma^2 \equiv -\frac{1}{3\,f''\left(0\right)} \label{gammadef}\\
	&&\,T_m \,=\, \eta_{\mu\nu}\,T_m^{\mu\nu} \label{emscaldef}
	\end{eqnarray}
	Derivation of the above differential equation from the trace of $ f\left(R\right) $ field equations have been given in Appendix \ref{ap:scalwav}. The source side of Eq. (\ref{newtracepow}) has an energy-momentum scalar $ \tilde{\Lambda} $ that is multiplied by a factor of $ \left(-g\right) $, similar to the energy-momentum tensor $ \tilde{\Lambda}^{\mu\nu} $ of Eq. (\ref{hfrakdyn}). It is to be noted that under the transformation (\ref{newtym}), the presence of the factor of $ \sqrt{-g} $ puts an $ \epsilon^{-1} $ dependence on the inverse length squared factor $ \gamma^2 $. This implies that for $ \epsilon\,\rightarrow\,0 $, the inverse length is essentially infinite, and the scalar field $ \mathfrak{R} $ cannot be excited. This also implies that the scalar field excitation is more for values of $ \epsilon $ close to unity, that is strong curvature along with high velocities in the strong field point particle formalism (\cite{Futamase2007}). This is known as the Chameleon mechanism or screening effect in the literature of $ f\left(R\right) $ and scalar-tensor theories.\\
	
	At this point one can impose the boundary conditions that have no radiative degrees of freedom (time dependent), tensor or scalar, coming from past null Minkowskian infinity or from any other parts of the space-time. This assumption implies that the objects in question are isolated objects at $ s\,=\,0 $, unaffected by other events in the universe. While this assumption gets rid of tensor/scalar radiation, one still has to consider non-radiative and time independent solutions of the homogeneous Klein-Gordon equation.\\
	\subsection{Homogeneous solution of the Klein-Gordon equation}\label{sec:kghomo}
	The time independent and homogeneous Klein-Gordon equation, being a second order differential equation of only spatial coordinates, is given by
	\begin{eqnarray}\label{homokg}
	\nabla\,\mathfrak{R}_{hom}\,-\,\frac{\gamma^2}{\epsilon}\,\,\mathfrak{R}_{hom}\,&=&\,0
	\end{eqnarray}
	where $ \nabla $ is the Laplace operator in Minkowski space-time, and the square root of the negative metric determinant was written in terms of the PN parameter, as seen after Eq. (\ref{scaledminkowski}). It is seen that the second term in the LHS of Eq. (\ref{homokg}) is of leading order in an $ \epsilon \to 0 $ series, and the system as a whole corresponds to the Klein-Gordon equation of a scalar field with infinite mass for $ \epsilon \to 0 $. Multiplying both sides of Eq. (\ref{homokg}) with $ \epsilon $, and taking the limit $ \epsilon \to 0 $, one obtains
	\begin{eqnarray}\label{newtorderhomokg}
	\mathfrak{R}_{hom} &=& 0
	\end{eqnarray}
	implying that the homogeneous solution cannot be excited at the leading order, or when the separation between a binary is large enough / their relative velocities are vanishing. However, at larger values of $ \epsilon $, the first term in the LHS of Eq. (\ref{homokg}) cannot be ignored, that is, Eq. (\ref{newtorderhomokg}) does not hold true. In spherical symmetry, for example, Eq. (\ref{homokg}) can be written as
	\begin{eqnarray}\label{kghomo}
	\frac{1}{r^2}\,\left[r^2\,\left(\mathfrak{R}_{hom}\right)_{,r}\right]_{,r}\,-\,\frac{\gamma^2}{\epsilon}\,\mathfrak{R}_{hom} &=& 0
	\end{eqnarray}
	which has a solution which is both regular at $ r\,=\,0 $ and $ r\,=\,\infty $, and is given by
	\begin{eqnarray}\label{riccihom}
	\mathfrak{R}_{hom} &=& \frac{C\,\,e^{-\frac{\gamma\,r}{\sqrt{\epsilon}}}}{r}
	\end{eqnarray}
	where $ C $ is a real constant of integration yet to be fixed. To fix such an arbitrary integration constant, it is necessary to look for the effect of the homogeneous solution on the solution of redefined goth metric perturbation, or $ \tilde{\mathfrak{h}}^{\mu\nu} $. Specifically, one needs to find the order of $ \tilde{\mathfrak{h}}^{\mu\nu} $ at which the effect of the scalar homogeneous solution first appear as a part of the effective source tensor $ t_{eff}^{\mu\nu} $.
	\subsection{The metric deviation in the presence of the homogeneous Ricci scalar}
	Since the leading order Ricci deviation is $ \mathcal{O}\left(1\right) $ due to the presence of a homogeneous solution $ \mathfrak{R}_{hom}$, it needs to be included in $ \tilde{\Lambda}^{\mu\nu} $ which sources $ \tilde{\mathfrak{h}}^{\mu\nu} $. The solution for $ \tilde{\mathfrak{h}}^{\mu\nu} $ within the boundary condition (\ref{boundcon}) is given by the following integral
	\begin{eqnarray}
	\tilde{\mathfrak{h}}^{\mu\nu}\left(s,\,x^i\right) &=& 4\,\left(-g\right)\,\int_{C\left(s,\,x^k;\,\epsilon\right)}\,d^3y\,\frac{\tilde{\Lambda}^{\mu\nu}\left(s\,-\,\epsilon\,\left|x^i\,-\,y^i\right|,\,y^i\right)}{\left|x^i\,-\,y^i\right|}\label{greenintegralh}
	\end{eqnarray}
	where the integral has its support in the past light cone of the event $ \left(s,\,x^i\right) $, as fixed by the boundary condition, and denoted by $ C\left(s,\,x^k;\,\epsilon\right) $. Concentrating only on the time-time component of the above, for example, under the temporal transformation (\ref{newtym}), $ \tilde{\mathfrak{h}}^{t\,t}\,=\,\epsilon^{-2}\,\tilde{\mathfrak{h}}^{s\,s} $ and $ \left(-g\right)\,=\,\epsilon^{-2} $. Also, it is to be noted that $ \lim\limits_{\epsilon\,\rightarrow\,0}d^3y\,\tilde{\Lambda}^{t\,t}\,=\,\left(\epsilon^6\,d^3X_L\right)\,\left(\epsilon^{-2}\,\tilde{\Lambda}^{s\,s}\right)\,\sim\,\mathcal{O}\left(\epsilon^4\right) $ under the scalings (\ref{newtym}) and (\ref{scalcoord}). Therefore, under the combined temporal and spatial coordinate scalings, the body-zone integral becomes
	\begin{eqnarray}
	\tilde{\mathfrak{h}}^{s\,s}\left(s,\,x^i\right) &=& 4\,\epsilon^4\,\sum_{L=1,\,2}\,\int_{\mathcal{B}_L}\,d^3X_L\,\frac{\tilde{\Lambda}^{s\,s}}{\left|Z^i_L\left(s\right)\,-\,\epsilon^2\,X^i_L\right|}
	\end{eqnarray}
	Now utilizing the definition of $ \tilde{\Lambda}^{\mu\nu} $ from Eq. (\ref{EMSDnetfR}), at the leading order, that is just considering the part of $ \tilde{\Lambda}^{\mu\nu} $ proportional to the classical matter energy-momentum-stress density tensor, one obtains
	\begin{eqnarray}
	\tilde{\mathfrak{h}}^{s\,s} &=& 4\,\epsilon^4\,\sum_{L=1,\,2}\,\int_{\mathcal{B}_L}\,d^3X_L\,\frac{T^{s\,s}\,\left(1\,-\,\epsilon\,f''\,\mathfrak{R}\right)}{\left|Z^i_L\,-\,\epsilon^2\,X^i_L\right|}
	\end{eqnarray}
	$ \epsilon $ series of the first part of the integral out of the two terms above lead to Eq. (\ref{fhssmul}), same as in GR. However, a more curious thing occurs when one considers the second term with $ \mathfrak{R}\,=\,\mathfrak{R}_{hom} $. That is the following
	\begin{eqnarray}
	\tilde{\mathfrak{h}}^{s\,s}_{extra} &=& -4\,\epsilon^5\,f''\,\sum_{L=1,\,2}\,\int_{\mathcal{B}_L}\,d^3X_L\,\frac{T^{s\,s}\,\mathfrak{R}_{hom}}{\left|Z^i_L\,-\,\epsilon^2\,X^i_L\right|}
	\end{eqnarray}
	Considering now the particular form of the homogeneous Ricci density (\ref{riccihom}), one obtains the following integral using the scaled coordinate (\ref{scalcoord})
	\begin{eqnarray}
	\tilde{\mathfrak{h}}^{s\,s}_{extra} &=& -4\,\epsilon^3\,f''\,\sum_{L=1,\,2}\,C_L\,\int_{\mathcal{B}_L}\,d^3X_L\,\frac{T^{s\,s}}{\left|Z^i_L\left(s\right)\,-\,\epsilon^2\,X^i_L\right|\,\left|X^i_L\right|}\,e^{-\gamma\,\epsilon^{\frac{3}{2}}\,\left|X^i_L\right|}
	\end{eqnarray}
	which seems to preceed the order at which the gothic metric deviation first appears, that is $ \mathcal{O}\left(\epsilon^4\right) $. However, under an $ \epsilon $ series about $ 0 $, the leading order form of the above becomes the following
	\begin{eqnarray}
	\tilde{\mathfrak{h}}^{s\,s}_{extra} &=& -4\,\epsilon^3\,f''\,\sum_{L=1,\,2}\,\frac{C_L}{\left|Z^i_L\right|}\,\int_{\mathcal{B}_L}\,d^3X_L\,\frac{T^{s\,s}}{\left|X^i_L\right|}\\
	\tilde{\mathfrak{h}}^{s\,s}_{extra} &=& -4\,\epsilon^3\,f''\sum_{L=1,\,2}\,\frac{C_L\,N_{L}}{\left|X^i_L\right|}
	\end{eqnarray}
	which is a Newtonian potential like term with negative one multipole moment of the source, denoted by $ N_L $. Therefore, the modified leading order metric deviation $ \tilde{\mathfrak{h}}^{s\,s} $ can be written in the following form
	\begin{eqnarray}
	\tilde{\mathfrak{h}}^{s\,s} &=& 4\,\epsilon^4\,\sum_{L=1,\,2}\,\left[\frac{\tilde{P}^s_L}{\left|Z^i_L\right|}\,+\,\mathcal{O}\left(\epsilon^2\right)\right]\\
	\tilde{P}^s_L &=& P^s_L\,-\,\lim\limits_{\epsilon\,\rightarrow\,0}\,\frac{f''\,C_L}{\epsilon}\,\int_{\mathcal{B}_L}\,\,d^3X_L\,\frac{T^{s\,s}}{\left|X^i_L\right|} \label{massdefdiv}
	\end{eqnarray}
	where $ P^s_L $ was defined in Eq. (\ref{monoss}), and is the classical definition of quasi-local mass at the leading / Newtonian order. From astrophysical observations, for example from solar system tests of gravity, a two-body system obeys Newton's law of gravity quite well, especially when their relative velocity divided by the speed of light tends to zero. In Newton's law, the only parameter that arises is the classical quasi-local mass ($ \frac{c^2\,P^s_L}{G} $ in non-geometrized units). More importantly, it is the very first / leading order parameter that arises in a PN expansion of GR, which has a very well defined $ \epsilon \to 0 $ limit, and which has been shown to hold rather well, even in situations like a binary neutron star system. In the current formalism, the classical qausi-local mass definition first appears at $ \mathcal{O}\left(\epsilon^4\right) $ for a PN expanded $ \mathfrak{h}^{s\,s} $ in GR. Noticing in Eq. (\ref{massdefdiv}), that $ \tilde{P}^s_L $ can be interpreted as a redefined mass for the leading order $ \tilde{\mathfrak{h}}^{s\,s} $, it is important to note that while the integral in the second term of Eq. (\ref{massdefdiv}) has a well defined limit for $ \epsilon\,\rightarrow\,0 $, the factor multiplying it does not. In fact, due to the $ \epsilon^{-1} $ nature of the factor, the second term as a whole blows up in the $ \epsilon\,\rightarrow\,0 $ limit. In order to avoid such a diverging behavior for the classical quasi-local mass definition for the small $ \epsilon $ limit, the only choice one has is the following	
	\begin{eqnarray}
	C_L &=& 0,
	\end{eqnarray}
	which makes the homogeneous Ricci density vanish at all orders of $ \epsilon $, leading to
	\begin{eqnarray}
	\tilde{P}^s_L &=& P^s_L
	\end{eqnarray}
	implying the leading order behavior of the new gothic metric deviation $ \tilde{\mathfrak{h}}^{\mu\nu} $ is the same as in GR. It is to be noted that such a limiting procedure to Newton's law is also used to fix the constant of integration for the homogeneous spherically symmetric solution of GR (the Schwarzschild mass). The vanishing of the homogeneous solution leaves one with just the inhomogeneous/particular solution, and Eq (\ref{newtracepow}) can now be solved in an order reduced method for the Ricci density $ \mathfrak{R} $ sourced by $ \tilde{\Lambda} $, as a Klein-Gordon equation with a scalar effective source respectively. The "non-linear terms" and the derivation of the above from Eq. (\ref{scalfield}) has been given in Appendix \ref{ap:scalwav}.\\
	
	Substantial amount of literature is dedicated to finding non-trivial (non Ricci flat) black hole solutions, that simultaneously solve the system of homogeneous partial differential equations (\ref{newtracepow}) and (\ref{hfrakdyn}). See for example constant curvature black hole space-times obtained in \cite{Dombriz2009}, solutions obtained from perturbing the Schwarzschild space-time in \cite{Nzioki:2014oaa}, and other non-trivial solutions in \cite{Nashed2020}. Also see \cite{Nashed2018} for spherically symmetric electro-vacuum solutions.\\
	
	Given the solution is being presented only around $ R\,=\,0 $, imply that at the first approximation, only the classical matter energy-momentum tensor trace $ T_m $ source the Ricci scalar density $ \mathfrak{R} $. The structure of the differential equation (\ref{newtracepow}), and the effective source in Eq. (\ref{effemtilde}), consists of a constant $ f''\left(0\right) $ with dimension of $ \left(length\right)^2 $, and an inverse length scale $ \gamma $. Expansion about $ R\,=\,0 $, and the assumption that the constant $ f''\left(0\right) $ has to be small compared to typical $ \left(\text{length}\right)^2 $ scales of the problem (from observations telling one that nature follows GR quite well) imply that in the RHS of Eq. (\ref{newtracepow}) the trace of the classical matter energy-momentum-stress density tensor affect the dynamics of $ \mathfrak{R} $ more strongly (that is, leading order in the PN expansion) than terms in the curly brackets of $ \tilde{\Lambda} $, in Eq. (\ref{effemtilde}), weighed by $ f''\left(0\right) $. The inhomogenoeus problem can hence be simplified into the following at the leading order
	\begin{eqnarray}\label{newtracesmall}
	\Box\,\mathfrak{R}\,+\,\frac{\gamma^2}{\epsilon}\,\mathfrak{R} &=& -8\,\pi\,\left(-g\right)\,\gamma^2\,T_m
	\end{eqnarray}
	\\
	Eqs. (\ref{hfrakdyn}) and (\ref{newtracepow}) form the complete set of equations required to obtain a PN sequence of solutions for $ \mathfrak{\tilde{h}}^{\mu\nu} $ and $ \mathfrak{R} $.
	\subsection{The Ricci scalar and its effect on the equations of motion}
	\label{ss:ricciandeom}
	The net Ricci solution, including the particular solution, can be written as a series about $ \epsilon\,\rightarrow\,0 $, and is given by (derivation in Appendix \ref{ap:kleingordon})
	\begin{eqnarray}\label{riccimultipole}
	\mathfrak{R}\left(t,\,x^i\right) &\approx& -8\,\pi\,\gamma^2\,\sum_{L,\,\ell,\,m}^{\ell\,=\,\infty}\,\epsilon^{2\,\ell\,+\,4}\,\frac{\left(\frac{\gamma^2}{\epsilon}\,-\,m^2\,\Omega^2\right)^{\frac{2\,\ell\,+\,1}{4}}}{2^{\ell\,+\,\frac{1}{2}}\,\Gamma\left(\ell\,+\,\frac{3}{2}\right)}\,\,\frac{K_{\ell\,+\,\frac{1}{2}}\left(\sqrt{\frac{\gamma^2}{\epsilon}\,-\,m^2\,\Omega^2}\,\left|Z^i_L\right|\right)}{\sqrt{\left|Z^i_L\right|}}\nonumber\\
	&&\times\,Y_{\ell\,m}\left(Z^\theta_L,\,Z^\phi_L\right)\,\mathfrak{M}_{L\,\ell\,m}\,\,e^{i\,m\,\left(\phi_0\,-\,\Omega\,t\right)}
	\end{eqnarray}
	where $ \Gamma\left(n\right) $ is the Euler-Gamma function, $ K_{\ell\,+\,\frac{1}{2}} $ is the modified Bessel function of the second kind, and $ Y_{\ell\,m} $ are spherical harmonic functions. $ \Omega $ corresponds to the quasi-adiabatic orbital angular frequency and $ \phi_0 $ corresponds to the phase lag between the orbital phase $ \Omega\,t $ and the corresponding phase of the Ricci scalar deviation. The phase lag is a result of the delay in response between the movement of the sources and the corresponding Ricci deviation generated by the accelerated motion of two ultra-compact object about each other. The scalar multipole moments $ \mathfrak{M}_{L\,\ell\,m} $ are defined as follows
	\begin{eqnarray}\label{mulscal}
	\mathfrak{M}_{L\,\ell\,m} &=& 2\,\pi\,\left(-1\right)^m\,N_{\ell\,-m}\,\int_{\mathcal{B}_L}\left|X^i_L\right|^{2+\ell}\,\sin \left(X^\theta_L\right)\nonumber\\
	&&\times\quad P_{\ell\,-m}\left(\cos X^\theta_L\right)\,\mathfrak{T}_m\left(\left|X^i_L\right|,\,X^\theta_L\right)d\left|X^i_L\right|\,dX^\theta_L\nonumber\\
	\\
	N_{\ell\,m} &=& \left(-1\right)^m\,\sqrt{\frac{\left(2\,\ell\,+\,1\right)}{4\,\pi}\,\frac{\left(\ell\,-\,m\right)!}{\left(\ell\,+\,m\right)!}}
	\end{eqnarray}
	where $ P_{\ell\,m}\left(X^\theta_L\right) $ are the associated Legendre polynomials. It is to be noted that $ \mathfrak{M}_{L\,0\,0} $, or the monopole term is the usual definition of mass, as in GR, as defined in Eq. (\ref{grmassdef}).\\
	
	It is seen that the Ricci scalar density scales as $ \epsilon^4 $ at the dominant order $ \left(\ell\,=\,0\right) $, and its dynamics is independent of the  dynamics of the modified gothic metric $ \mathfrak{\tilde{h}}^{\mu\nu} $. It is however, for the sake of obtaining observational consequences of the current theory, more interesting to look at the opposite, that is the effect of the Ricci scalar deviation on the dynamics of $ \mathfrak{\tilde{h}}^{\mu\nu} $, and consequently, its gauge invariant contribution to the modified equations of motion of the binary system.\\
	
	The equations of motion can be found by using the modified conservation law (\ref{effemcons}) alongwith the definitions of (\ref{momevol}) and (\ref{vevol}) with $ \Lambda^{\mu\nu} $ replaced by $ \tilde{\Lambda}^{\mu\nu} $. Since the equations of motion involve a vector surface integral over the body zone boundary, on which $ T_m^{\mu\nu} $ vanishes (the classical matter source being a compact one), and consequently so does the effect of the first term in the RHS of Eq. (\ref{teffrel}) in the equations of motion. A vector surface integral over a sphere also vanishes if the integrand is spherically symmetric, which $ {}_{(4)}\mathfrak{R} $ is, as seen in (\ref{ricciinhom}). Therefore, the terms that might contribute to a deviation from GR at the leading order are the space-space components of the effective energy-momentum tensor $ t^{\mu\nu}_{eff} $, given by
	\begin{eqnarray}
	&&t^{i\,j}_{eff} = \frac{\left(-g\right)^{-2}}{192\,\pi\,\gamma^4}\,\left[2\,\left(\eta^{i\,k}\,\eta^{j\,l}\,+\,\eta^{i\,l}\,\eta^{j\,k}\,-\,\eta^{i\,j}\,\eta^{k\,l}\right)\,\mathfrak{R}\,\partial_k\mathfrak{R}\,\partial_l\tilde{\mathfrak{h}}\right.\nonumber\\
	&&\left.\,+\,\left(\eta^{i\,k}\,\eta^{j\,l}\,-\,\frac{1}{2}\,\eta^{i\,j}\,\eta^{k\,l}\right)\,\left(4\,\partial_k\mathfrak{R}\,\partial_l\mathfrak{R}\,+\,\mathfrak{R}^2\,\partial_k\tilde{\mathfrak{h}}\,\partial_l\tilde{\mathfrak{h}}\right)\right]\label{lambdasix}
	\end{eqnarray}
	owing to the fact that at the leading order, both the metric density deviation trace $ \tilde{\mathfrak{h}} $ and the Ricci deviation density $ \mathfrak{R} $ are time independent, and $ \left|\eta^{s\,s}\right|\,\sim\,\epsilon^2 $.\\
	
	One notices that the leading and sub-leading order Ricci density scale as $ \mathcal{O}\left(\epsilon^4\right) $, and correspondingly, the deviation from GR in the equations of motion of the binary appears at $ \mathcal{O}\left(\epsilon^{12}\right) $, which leads to the modification of the 6 PN equations of motion, which are then given by the following
	\begin{eqnarray}
	M_1\,\frac{dv_1^i}{ds} &=& F^i_{Newtonian} \,+\, \epsilon^2\,F^i_{1PN,GR} \,+\, \epsilon^4\,F^i_{2PN,GR} \,+\, \epsilon^5\,F^i_{2.5PN,GR} \nonumber\\
	&&\,+\,F^i_{3PN,GR}\,+\,...\,+\,\epsilon^{12}\,\left(F^i_{6PN,GR}\,-\,\frac{32\,\pi\,M_1\,M_2^3}{9\,r_{1\,2}^4}\,\,e^{-\frac{2\,\gamma\,r_{1\,2}}{\sqrt{\epsilon}}}\,n^i\right)  \nonumber\\ \label{fReom}
	&& \,+\, \mathcal{O}\left(\epsilon^{13}\right)
	\end{eqnarray}\\
	The derivation of the above is given in Appendix \ref{ap:kleingordon}. $ F_{N}^i $ and $ F_{kPN,GR}^i $ are the attractive Newtonian and $ k^{th}\, \left(k\,=\,1,...,6\right) $ PN forces in GR, respectively, on the first body zone.\\

	The leading modification at 6 PN to the equations of motion is seen to be repulsive at very short scales, and for very high velocities and compactness, given the convention chosen in this paper regarding the direction of the unit normal $ n^i $, which points from the first body zone towards the COM of the second source. The leading modification is dependent on a length scale that acts like a new universal constant (if nature follows such a quadratic theory) and  is $ \sim\,\gamma^{-1} $. It is interesting to note that in the 6 PN extra force term, one can cancel $ M_1 $ from both sides leading to only $ M_2^3 $ remaining as a product in the entire RHS of Eq. (\ref{fReom}). This establishes that the equivalence/effacing nature of two ultra-compact sources in a binary orbit around each other till the 6 PN order in polynomial $ f\left(R\right) $ theories of gravity with $ f''\left(0\right)\,<\,0 $ remains almost the same as in GR.
	\section{Discussions and Conclusions}
	\label{s:DisCo}
	In this article the PN formalism of GR was reviewed using the surface integral approach \cite{Einstein1949OnTM}, using the specially scaled co-moving coordinate system for the internal problem \cite{Futamase2007}. Using the strong field point particle mechanism developed for the PN expansion of GR in \cite{Futamase2007}, a quadratic $ f\left(R\right) $ model of gravity was PN expanded in the conformal-harmonic gauge. In doing so, it was clearly established that no deviation from GR occurs in the quadratic $ f\left(R\right) $ theory till the sixth post Newtonian order, thus automatically also taking care of the effacing nature of sources in a binary orbit.\\
	
	A scaled co-moving coordinate system, and the Einstein-Infeld approach \cite{Einstein1949OnTM} was used in order to solve for the two body equations of motion in $ f\left(R\right) $ theories of gravity. A conformal scaling of the old definition of the gothic metric of GR was found to be more suited to get possible observational consequences out of a theory that is highly non-linear and fourth derivative of the metric tensor. Under the conformal scaling, the field equations resemble the familiar equations of the Landau-Lifshitz formulation of GR, with an extra 'effective' source term that is the manifestation of the extra scalar degree of freedom in $ f\left(R\right) $ theories.\\
	
	More notably, a dynamical scalar variable for solving the combined tensor+scalar system of second order partial differential equations was found: the Ricci scalar density $ \mathfrak{R} $. This particular redefinition frames the tensor differential equations and the scalar Klein-Gordon equation in a similar footing; that is expanded in a series of $ \frac{f''}{\sqrt{-g}} $, that is, $ \epsilon\,f'' $. Coupling the scalar variable, the Ricci scalar, with the volume element $ \sqrt{-g} $ in the harmonic gauge of Eq. (\ref{conlor}) leads to a system of differential equations that can be solved perturbatively in terms of the PN parameter $ \epsilon $, that smoothly evolves from zero to unity. The redefinition allows one to ignore the Klein-Gordon homogeneous solution; and to sieve through non-linear terms of $ \tilde{\mathfrak{h}}^{\mu\nu} $ and $ \mathfrak{R} $ (and their first derivatives) in the effective energy-momentum tensor $ t^{\mu\nu}_{eff} $, and obtain terms that are leading with regards to the parameter $ \epsilon $, as well as group them in terms of sub-leadingness.\\
	
	It was shown that the homogeneous Ricci scalar density has to vanish if the definition of mass at the Newtonian level is well defined. This then leads to the inhomogeneous Ricci scalar density solution appearing at $ \mathcal{O}\left(\epsilon^4\right) $, and sourcing the metric density tensor deviation $ \tilde{\mathfrak{h}}^{\mu\nu} $ through the effective source term $ t^{\mu\nu}_{eff} $ of $ f\left(R\right) $ theories of gravity.\\
	
	In this study, the change in the equations of motion of two extended (but compact objects) from GR in the quadratic Ricci model till the $ 6^{th} $ PN order were obtained. The equations of motion till 5.5 PN order remain unchanged compared to GR. A modification of the equations of motion from the GR form, through the appearance of a repulsive Yukawa like force at length scales comparable to $ \gamma^{-1} $ at the $ 6^{th} $ PN order was observed. However, the modified force is time independent, conservative, and dependent only on the usual definition of mass \& the orbital separation. A universal constant $ f''\left(0\right) $, other than the gravitational constant (put to unity in Geometrized units), was defined, similar to other works in the literature. The constant has dimensions of $ \left(length\right)^2 $, and a corresponding inverse length scale $ \gamma $. \\
	
	The above analysis also brings to light a possibly new class of solutions of $ f\left(R\right) $ theories of gravity, and shows that the appearance of structure in the equations of motion in $ f\left(R\right) $ theories are exactly the same as in GR, to a very high PN accuracy. At the leading order of deviation from GR, no extra structure comes into the equations of motion due a quadratic change in the action. The appearance of structure at high PN orders till 6 PN is still solely a GR effect. This is the strong equivalence principle of $ f\left(R\right) $ theories in action. The current study establishes the proposition that both the weak and strong equivalence principles hold for $ f\left(R\right) $ theories of gravity, and tests for the validity of GR through tests of the equivalence principle will not at all distinguish between the two theories. Hence, the equivalence, and consequently, the effacing principle is not unique to GR.
	\section*{Acknowledgement}
	The author would like to thank Bruce Allen and Badri Krishnan for a significant number of ideas and clarifications throughout the project. The author thanks Pierre Mourier for important comments on earlier drafts. The author would like to thank Yousuke Itoh, and the CBC group of MPI for Gravitational Physics, especially Andrey Shoom, for important comments and questions. The author would like to thank Chowhan Tanmay Tushar and Amanda Fajardo Garcia for some key inputs to the calculations of the Landau-Lifshitz formalism of GR.\\
	
	The technical computing system Mathematica, and some of the symbolic tensor manipulation packages supported by it, namely $ xAct $ \cite{Mart_n_Garc_a_2008} was used for most of the calculations in the study.
	\appendix
	\section{Taming the scalar wave equation for the extra massive degree of freedom of $ f\left(R\right) $}
	\label{ap:scalwav}
	The scalar wave dynamics of Eq. (\ref{scalfield}) is non-linear, as seen in terms like $ f' R $. By using the definition of the Laplace-Beltrami operator, one may write (\ref{hfrakdyn}) as
	\begin{eqnarray}
	\frac{3}{\sqrt{-g}}\,\left(\sqrt{-g}\,g^{\mu\nu}\,f'_{\,,\,\nu}\right)_{\,,\,\mu}\,+\,\left(f'\,R\,-\,2\,f\right) &=& 8\,\pi\,T_m\nonumber\\
	\end{eqnarray}
	Noticing the old gothic metric definition $ \left(\sqrt{-g}\,g^{\mu\nu}\right) $ in the first term, one can conformally transform the old definition to the new definition using Eq. (\ref{newgoth}) and perform the perturbation in Eq. (\ref{tenspert}), after which the above equations become
	\begin{eqnarray}
	&&-\frac{3}{\sqrt{-g}\,f'}\,\left[\Box f'\,-\,\mathfrak{\tilde{h}}^{\mu\nu}\,f'_{\,,\,\mu\nu}\,-\,\frac{1}{f'}\,\left\{\left(f'\right)^{\,,\,\mu}\,f'_{\,,\,\mu}\right.\right.\nonumber\\
	&&\left.\left.+\,\mathfrak{\tilde{h}}^{\mu\nu}\,f'_{\,,\,\mu}f'_{\,,\,\nu}\right\}\right]\,+\,\left(f'\,R\,-\,2\,f\right) = 8\,\pi\,T_m
	\end{eqnarray}
	The second and the fourth term inside the square brackets can be transformed, by chain rule, into total derivative terms, and terms where the Lorenz gauge conditions can be imposed. Noting that boundary dependent terms can be discarded to obtain gauge invariant objects, one obtains
	\begin{eqnarray}
	\Box f'\,+\,\sqrt{-g}\left[\frac{1}{3}\,\left(f'\right)^2\,R\,-\,\frac{2}{3}\,f\,f'\,-\,\frac{8\,\pi}{3}\,f'\,T_m\right] &=& 0\nonumber\\
	\end{eqnarray}
	Now the particular forms for $ f\left(R\right) $ and $ f'\left(R\right) $ can be imposed, according to Eqs. (\ref{fRform}) and (\ref{fpRform}), as well as the dynamical variable definition for the Ricci scalar density $ \mathfrak{R} $ in Eq. (\ref{scalpert}), which leads to 
	\begin{eqnarray}\label{kgordonwithsource}
	&&\Box\,\mathfrak{R}\,+\,\sqrt{-g}\,\gamma^2\,\mathfrak{R} = -8\,\pi\,\left(-g\right)\,\gamma^2\,\tilde{\Lambda}\\
	&&\tilde{\Lambda} = \,T_m\,+\,f''\left(0\right)\,\left\{\frac{T_m\,\mathfrak{R}}{3\,\sqrt{-g}}+\,\frac{1}{8\,\pi\,\left(-g\right)}\left(\frac{\mathfrak{R}^2}{3}\,-\,\partial_\mu\mathfrak{R}\,\partial^\mu\tilde{\mathfrak{h}}\,-\,\frac{1}{2}\,\mathfrak{R}\,\partial_{\mu}\tilde{\mathfrak{h}}^{\alpha\beta}\,\partial^{\mu}\tilde{\mathfrak{h}}_{\alpha\beta}\right.\right.\nonumber\\
	&&\left.\left.\,-\,\frac{1}{4}\,\mathfrak{R}\,\partial_{\mu}\tilde{\mathfrak{h}}\,\partial^\mu\tilde{\mathfrak{h}}\,-\,\frac{1}{2}\,\mathfrak{R}\,\Box\tilde{\mathfrak{h}}\right)\right\}\,+\,\mathcal{O}\left(\left[f''\left(0\right)\right]^2\right)\\
	&&\gamma^2 \equiv -\frac{1}{3\,f''\left(0\right)} \\
	&&\,T_m \,=\, \eta_{\mu\nu}\,T_m^{\mu\nu}
	\end{eqnarray}
	where the argument of $ \mathcal{O} $ in the above have been made dimensionless. Since in the scaled time, inverse of the square root of the determinant of the metric scale as $ \epsilon $, corrections to the effective energy-momentum scalar $ \tilde{\Lambda} $ from $ T_m $ at the next-to-leading order will be of $ \mathcal{O}\left[\epsilon^2\,\frac{f''\left(0\right)}{\left(\text{length}\right)^2}\right] $, with $ \left(\text{length}\right) $ being typical length scales of the binary problem.
	\section{Expanding the solution of the order reduced Klein-Gordon equation in orders of $ \epsilon $}
	\label{ap:kleingordon}
	The net solution to the  Klein-Gordon problem (\ref{newtracesmall}) with the boundary condition choice of no incoming scalar radiation from past null Minkowskian infinity is then given as follows
	\begin{eqnarray}\label{greenint}
	\mathfrak{R}\left(x^\mu\right) &=& -8\,\pi\,\gamma^2\,\epsilon^{-2}\,\int\,d^4y\, G_{\tilde{\gamma}}\left(x^\mu,\,y^{\mu}\right)\,T_m\left(y^{\mu}\right)\\
	\tilde{\gamma}^2 &\equiv& \frac{\gamma^2}{\epsilon}
	\end{eqnarray}
	where the factor of $ \epsilon^{-2} $ arises because the metric determinant $ \left(-g\right) $ multiplying the RHS of Eq. (\ref{newtracesmall}), under the transformation (\ref{newtym}), scale as $ \epsilon^{-2} $, as seen after Eq. (\ref{scaledminkowski}). $ G_{\tilde{\gamma}}\left(x^\mu,\,y^\mu\right) $ is the retarded Green's function of the Klein-Gordon equation, as was given in \cite{Berry2011} with the $ \left(+\,,\,-\,,\,-\,,\,-\right) $ metric signature as
	\begin{eqnarray}\label{kgreen}
	G_{\tilde{\gamma}}\left(t,\,q;\,x^i,\,y^i\right) &=& \int_{-\infty,\,\tilde{\gamma}}^{-\tilde{\gamma},\,\infty} \frac{d\,\omega}{2\,\pi}\,e^{-i\,\omega\left(t-q\right)}\,\frac{e^{i\,\sqrt{\omega^2-\tilde{\gamma}^2}\,\left|x^i-y^i\right|}}{4\,\pi\,\left|x^i-y^i\right|} \nonumber\\
	&&+\,\int_{-\tilde{\gamma}}^{\tilde{\gamma}}\,\frac{d\omega}{2\,\pi}\,e^{-i\,\omega\,\left(t-q\right)}\,\frac{e^{-\,\sqrt{\tilde{\gamma}^2-\omega^2}\,\left|x^i-y^i\right|}}{4\,\pi\,\left|x^i-y^i\right|}\nonumber\\
	\end{eqnarray}
	The notation $ \int_{-\infty,\,\tilde{\gamma}}^{-\tilde{\gamma},\,\infty} $ involve two integrals, one from $ -\infty $ to $ -\tilde{\gamma} $, and the other from $ \tilde{\gamma} $ to $ \infty $. It is to be noted that the three integrals whose domains encompass all of $ \omega $ space must be evaluated simultaneously in order for the solution to converge. $ T_m\left(q,\,y^i\right) $, the trace of the classical matter energy-momentum tensor, which being a scalar, does not transform under any of the coordinate scalings.\\
	
	It is to be noted that the PN parameter $ \epsilon $ appears explicitly in the Green's function, both as an argument of the exponential function, and as one (or both) of the limits of the two $ \omega $ space integrals at hand. Considering the first integral, that is
	\begin{eqnarray}
	I_1 &=& \int_{-\infty,\,\tilde{\gamma}}^{-\tilde{\gamma},\,\infty} \frac{d\,\omega}{2\,\pi}\,e^{-i\,\omega\left(t-q\right)}\,\frac{e^{i\,\sqrt{\omega^2-\tilde{\gamma}^2}\,\left|x^i-y^i\right|}}{4\,\pi\,\left|x^i-y^i\right|}
	\end{eqnarray}
	In the Newtonian limit, or $ \epsilon \to 0 $, $ \tilde{\gamma} $ diverges. But since it appears in the argument of the oscillatory function, that is $ e^{i\,\sqrt{\omega^2-\tilde{\gamma}^2}\,\left|x^i-y^i\right|} $, the Green's function itself does not diverge and remains finite in $ \left[-1,1\right] $ as long as $ \omega^2>\tilde{\gamma}^2 $. Although it becomes a highly oscillating function. At the same time the limits of the integral are replaced with $ \pm\tilde{\gamma} \to \pm\infty $, effectively shrinking the domain of integration to a vanishing value (implying that the lower and upper limits of the two integrals coincide for $ \epsilon \to 0 $). This leads to a vanishing value of $ I_1 $ for $ \epsilon \to 0 $. The second integral, that is when $ \tilde{\gamma}^2>\omega^2 $, which is the following
	\begin{eqnarray}
	I_2 &=& \int_{-\tilde{\gamma}}^{\tilde{\gamma}}\,\frac{d\omega}{2\,\pi}\,e^{-i\,\omega\,\left(t-q\right)}\,\frac{e^{-\,\sqrt{\tilde{\gamma}^2-\omega^2}\,\left|x^i-y^i\right|}}{4\,\pi\,\left|x^i-y^i\right|},
	\end{eqnarray}
	The limit $ \epsilon \to 0 $ leads to the domain of integration covering the whole of Real number space. However, the argument in the decaying function $ e^{-\,\sqrt{\tilde{\gamma}^2-\omega^2}\,\left|x^i-y^i\right|} $ blows up leading to the exponential vanishing, and hence leading to $ I_2 \to 0 $ for $ \epsilon \to 0 $.\\
	
	Due to the the coordinate scalings in Eq. (\ref{scalcoord}), the infinitesimal 4-volume element $ d^4y $ transform from the asymptotic observer's frame to either of the body zones $ \mathcal{B}_L $ in the following manner
	\begin{eqnarray}
	d^4y &\equiv& dt\,\wedge\,d^3y \\
	&\rightarrow& \epsilon^{-1}\,ds\,\wedge\,\epsilon^6\,\,d^3X_L \label{volelbod}
	\end{eqnarray}
	where $ \wedge $ denotes the wedge product between 1-form dt and 3-form $ d^3y $. In the current operational context $ \wedge $ is effectively scalar multiplication.\\
	
	Substituting Eq. (\ref{kgreen}) in Eq. (\ref{greenint}), and transforming into the body zone coordinates by substituting Eq. (\ref{volelbod}) as the infinitesimal covariant volume element in Eq. (\ref{greenint}), one obtains the following integral for the particular solution $ \mathfrak{R}_{part} $ for the inhomogeneous Klein-Gordon equation
	\begin{eqnarray}\label{rstep1}
	\mathfrak{R}_{part}\left(t,\,x^i\right) &=& -\epsilon^3\,\,8\,\pi\,\gamma^2\,\int_{-\infty}^{\infty}\,ds'\,\int_{\mathcal{B}_L}\,d^3X_L\,\,\int_{-\infty,\,\tilde{\gamma}}^{-\tilde{\gamma},\,\infty}\,\frac{d\omega}{2\,\pi}\,e^{-i\,\omega\,\left(t-s'/\epsilon\right)}\nonumber\\
	&& \times\quad\,\frac{e^{i\,\sqrt{\omega^2-\tilde{\gamma}^2}\,\left|Z^i_L-\epsilon^2\,X^i_L\right|}}{4\,\pi\,\left|Z^i_L-\epsilon^2\,X^i_L\right|}\,\left[\epsilon^4\,T_m\left(s'/\epsilon,\,X_L^i\right)\right]\nonumber\\
	&&-\,\epsilon^3\,\quad\,8\,\pi\,\gamma^2\,\int_{-\infty}^{\infty}\,ds'\,\int_{\mathcal{B}_L}\,d^3X_L\,\int_{-\tilde{\gamma}}^{\tilde{\gamma}}\,\frac{d\omega}{2\,\pi}\,e^{-i\,\omega\,\left(t-s'/\epsilon\right)}\nonumber\\
	&&\times\quad\,\frac{e^{-\sqrt{\tilde{\gamma}^2-\omega^2}\,\left|Z_L^i-\epsilon^2\,X^i_L\right|}}{4\,\pi\,\left|Z_L^i-\epsilon^2\,X^i_L\right|}\,\left[\epsilon^4\,T_m\left(s'/\epsilon,\,X^i_L\right)\right]\nonumber\\
	\end{eqnarray}
	
	In the body zones that surround each of the objects, the energy-momentum tensor of classical matter $ \left(T_{\mathcal{B}_L}\,=\,\lim\limits_{\epsilon\,\rightarrow\,0}\epsilon^4\,T_m\right) $ will be assumed to have a quasi-stationary/adiabatic initial condition for solving the relaxed system of equations, in the absence of any other time dependent driving force inside the compact object, and is given by
	\begin{eqnarray}\label{sauceass}
	T_{\mathcal{B}_L}\left(X^i\right) &=&  \sum_{n=-\infty}^{\infty}\,\mathfrak{T}_n\left(\left|X^i_L\right|,X^\theta_L\right)\,e^{i\,n\,\mathcal{X}}\\
	\mathcal{X}\left(s/\epsilon,\,X_L^\phi\right) &\equiv& \mathcal{X}\,=\,X^\phi_L\,-\,\Omega\,s/\epsilon\,+\,\phi_0 
	\end{eqnarray}
	where an axial+time symmetry was assumed for the initial condition, such that the time dependence of the source body in the body zone coordinate system repeats after every $ T\,=\,\frac{2\,\pi}{\Omega} $, with a constant phase parameter $ \phi_0 $. The weighing factors $ \mathfrak{T}_n $ are coefficients in the series expansion of the trace $ T_m \left(\text{or } T_{\mathcal{B_L}}\right) $ using stationary functions $ e^{i\,n\,\mathcal{X}} $, and are functions of the radial and azimuthal coordinates. Eq. (\ref{sauceass}) physically implies that the variations in the energy-momentum tensor sourcing $ \mathfrak{R} $ in the body zone of the first object is purely generated by the effect of the motion of the second object around it. The body zone coordinates $ \left(X^i_L\right) $ were defined in Eq. (\ref{scalcoord}), in which $ \left|X^i_L\right| $ is the distance from the center of mass of the body $ L $ to any point in the body zone coordinates, as viewed in the respective body zones. The choice in Eq. (\ref{sauceass}) has a simplifying effect on the subsequent calculations and is justified by the adiabatic and stationarity in the co-moving frame approximations, as found in the literature on PN expansions. \\
	
	Assuming quasi-periodicity for the source tensor, as in Eq. (\ref{sauceass}), one can substitute it into Eq. (\ref{rstep1}), to obtain
	\begin{eqnarray}
	\mathfrak{R}_{part}\left(t,\,x^i\right) &=& -\epsilon^3\,\,\frac{8\,\pi\,\gamma^2}{2\,\pi}\,\sum_{n=-\infty}^{\infty}\,\int_{\mathcal{B}_L}\,d^3X_L\,\left(\int_{-\infty,\,\tilde{\gamma}}^{-\tilde{\gamma},\,\infty}\,d\omega\,e^{-i\,\omega\,t}\right)\,e^{i\,n\,\left(X^\phi_L+\phi_0\right)}\nonumber\\
	&&\times\quad\,\frac{e^{i\,\sqrt{\omega^2-\tilde{\gamma}^2}\,\left|Z^i_L-\epsilon^2\,X^i_L\right|}}{4\,\pi\,\left|Z^i_L-\epsilon^2\,X^i_L\right|}\,\mathfrak{T}_n\left(\left|X^i_L\right|,X^\theta_L\right)\,\int_{-\infty}^{\infty}\,ds'\,e^{i\,\left(\omega-n\,\Omega\right)\,s'/\epsilon}\nonumber\\
	&&\,-\,\epsilon^3\quad\,\frac{8\,\pi\,\gamma^2}{2\,\pi}\,\sum_{n=-\infty}^{\infty}\,\int_{\mathcal{B}_L}\,d^3X_L\,\left(\int_{-\tilde{\gamma}}^{\tilde{\gamma}}\,d\omega\,e^{-i\,\omega\,t}\right)\,e^{i\,n\,\left(X^\phi_L+\phi_0\right)}\nonumber\\
	&&\times\quad\,\frac{e^{-\,\sqrt{\tilde{\gamma}^2-\omega^2}\,\left|Z^i_L-\epsilon^2\,X^i_L\right|}}{4\,\pi\,\left|Z^i_L-\epsilon^2\,X^i_L\right|}\,\mathfrak{T}_n\left(\left|X^i_L\right|,X^\theta_L\right)\,\int_{-\infty}^{\infty}\,ds'\,e^{i\,\left(\omega-n\,\Omega\right)\,s'/\epsilon}
	\end{eqnarray}
	Transforming
	\begin{eqnarray}
	s' \rightarrow \epsilon\,s
	\end{eqnarray}
	and using the integral representation of the delta function,
	\begin{eqnarray}
	\int_{-\infty}^{\infty}\,e^{i\,k\,\left(x-x'\right)}\,dk &=& 2\,\pi\,\delta\left(x-x'\right),
	\end{eqnarray}
	one obtains
	\begin{eqnarray}\label{riccisum}
	\mathfrak{R}_{part}\left(t,\,x^i\right) &=& -\epsilon^4\,\,8\,\pi\,\gamma^2\,\sum_{n=-\infty}^{\infty}\,\int_{\mathcal{B}_L}\,d^3X_L\,\frac{e^{i\,\sqrt{n^2\,\Omega^2-\tilde{\gamma}^2}\,\left|Z^i_L-\epsilon^2\,X^i_L\right|}}{4\,\pi\,\left|Z^i_L-\epsilon^2\,X^i_L\right|}\nonumber\\
	&&\times\quad\,\mathfrak{T}_n\left(\left|X^i_L\right|,X^\theta_L\right)\,\times\,e^{-i\,n\,\left(\Omega\,t\,-\,X^\phi_L\,-\,\phi_0\right)}\\
	&&-\epsilon^4\,\,8\,\pi\,\gamma^2\,\sum_{n=-\infty}^{\infty}\,\int_{\mathcal{B}_L}\,d^3X_L\,\frac{e^{-\,\sqrt{\tilde{\gamma}^2-n^2\,\Omega^2}\,\left|Z^i_L-\epsilon^2\,X^i_L\right|}}{4\,\pi\,\left|Z^i_L-\epsilon^2\,X^i_L\right|}\nonumber\\
	&&\times\quad\,\mathfrak{T}_n\left(\left|X^i_L\right|,X^\theta_L\right)\,e^{-i\,n\,\left(\Omega\,t\,-\,X^\phi_L\,-\,\phi_0\right)}
	\end{eqnarray}
	
	The Green's function (\ref{kgreen}) can be written as an infinite sum of spherical harmonic functions, that are weighed by functions of the radial coordinates $ \left|Z^i_L\right| $ and $ \left|X^i_L\right| $, in the following way for $ \left|Z^i_L\right|\,>\,\left|X^i_L\right| $
	\begin{eqnarray}
	\frac{e^{i\,k\,\left|X^i\,-\,Z^i\right|}}{4\,\pi\,\left|X^i\,-\,Z^i\right|} &=& i\,k\,\sum_{\ell,\,m}j_\ell\left(k\,\left|X^i\right|\right)\,h_\ell^{(1)}\left(k\,\left|Z^i\right|\right)\,Y^*_{\ell\,m}\left(X^\theta,\,X^\phi\right)\,Y_{\ell\,m}\left(Z^\theta,\,Z^\phi\right) \nonumber\\\label{greenexpupper}\\
	\frac{e^{-\,k\,\left|X^i\,-\,Z^i\right|}}{4\,\pi\,\left|X^i\,-\,Z^i\right|} &=& \sum_{\ell,\,m}\,\frac{I_{\ell+\frac{1}{2}}\left(k\,\left|X^i\right|\right)\,K_{\ell+\frac{1}{2}}\,\left(k\,\left|Z^i\right|\right)}{\sqrt{\left|X^i\right|\,\left|Z^i\right|}}\,Y^*_{\ell\,m}\left(X^\theta,\,X^\phi\right)\,Y_{\ell\,m}\left(Z^\theta,\,Z^\phi\right)\nonumber\\\label{greenexplower}
	Z^i_L &\equiv& \frac{x^i\,-\,z^i_L\left(s\right)}{\epsilon^2}\,;\quad\,L\,=\,1,\,2.
	\end{eqnarray}
	The various functions appearing above are as follows
	\begin{itemize}
		\item $ j_\ell $: Spherical Bessel function of first kind.
		\item $ h_\ell^{(1)} $: Spherical Bessel function of third kind.
		\item $ I_{\ell\,+\,\frac{1}{2}} $: Modified Bessel function of the first kind.
		\item $ K_{\ell\,+\,\frac{1}{2}} $: Modified Bessel function of the second kind.
		\item $ Y_{\ell\,m},\,Y^*_{\ell\,m} $: Spherical harmonic functions and their complex conjugates.
	\end{itemize}
	
	After using Eqs. (\ref{greenexpupper}) and (\ref{greenexplower}) in the above, one obtains for $ \left|Z^i_L\right|\,>\,\left|X^i_L\right| $
	\begin{widetext}
		\begin{eqnarray}\label{riccipostexp}
		\mathfrak{R}_{part}\left(t,\,x^i\right) &=& -\epsilon^4\,\,8\,\pi\,i\,\gamma^2\,\sum_{n,\,\ell,\,m}^{\left|n\right|\,>\,\floor*{\frac{\tilde{\gamma}}{\Omega}}}\,\sqrt{n^2\,\Omega^2-\tilde{\gamma}^2}\,\int_{\mathcal{B}_L}\,d^3X_L\,j_\ell\left(\epsilon^2\,\sqrt{n^2\,\Omega^2-\tilde{\gamma}^2}\,\left|X_L^i\right|\right)\nonumber\\
		&&\times\,\,h_\ell^{(1)}\left(\sqrt{n^2\,\Omega^2-\tilde{\gamma}^2}\,\left|Z^i\right|\right)\quad\,Y^*_{\ell\,m}\left(X_L^\theta,\,X_L^\phi\right)\,Y_{\ell\,m}\left(Z^\theta,\,Z^\phi\right)\,\mathfrak{T}_n\left(\left|X_L^i\right|,\,X_L^\theta\right)\,e^{i\,n\,\left(X_L^\phi+\phi_0-\Omega\,t\right)}\nonumber\\
		&&-\,\epsilon^3\,\quad\,\,8\,\pi\,\gamma^2\,\sum_{n,\,\ell,\,m}^{\left|n\right|\,<\,\floor*{\frac{\tilde{\gamma}}{\Omega}}}\,\int_{\mathcal{B}_L}\,d^3X_L\,\frac{I_{\ell+\frac{1}{2}}\left(\epsilon^2\,\sqrt{\tilde{\gamma}^2-n^2\,\Omega^2}\,\left|X_L^i\right|\right)}{\sqrt{\left|X_L^i\right|}}\,\frac{K_{\ell+\frac{1}{2}}\left(\sqrt{\tilde{\gamma}^2-n^2\,\Omega^2}\,\left|Z^i\right|\right)}{\sqrt{\left|Z^i\right|}}\nonumber\\
		&&\times\quad\,Y^*_{\ell\,m}\left(X_L^\theta,\,X_L^\phi\right)\,Y_{\ell\,m}\left(Z^\theta,\,Z^\phi\right)\,\mathfrak{T}_n\left(\left|X_L^i\right|,\,X_L^\theta\right)\,e^{i\,n\,\left(X_L^\phi+\phi_0-\Omega\,t\right)}\nonumber\\
		\end{eqnarray}
	\end{widetext}
	one can substitute the following properties of spherical harmonic functions in Eq. (\ref{riccipostexp})
	\begin{eqnarray}
	&&Y^*_{\ell\,m} = \left(-1\right)^m\,Y_{\ell\,-m}\\
	&&Y_{\ell\,-m} = N_{\ell\,-m}\,P_{\ell\,-m}\,\left(\cos\,X^\theta_L\right)\,e^{-i\,m\,X^\phi}\\
	&&\int_{0}^{2\,\pi}\,dX^\phi_L\,Y^*_{\ell\,m}\,e^{i\,n\,X^\phi} = 2\,\pi\,\left(-1\right)^m\,N_{\ell\,-m}\,P_{\ell\,-m}\left(X^\theta_L\right)\,\delta_{m\,n}\nonumber\\
	\label{intprop}\\
	&&N_{\ell\,m} = \left(-1\right)^m\,\sqrt{\frac{\left(2\,\ell\,+\,1\right)}{4\,\pi}\,\frac{\left(\ell\,-\,m\right)!}{\left(\ell\,+\,m\right)!}}
	\end{eqnarray} 
	where $ P_{\ell\,m}\left(X^\theta_L\right) $ are the associated Legendre polynomials and $ \delta_{m\,n} $ is the Kronecker delta distribution. Since $ m $ takes values between $ -\ell $ to $ \ell $, application of the Kronecker delta leads to the summation on $ \ell $ in the first term of Eq. (\ref{riccipostexp}) going from the lower integral part of $ \frac{\tilde{\gamma}}{\Omega} $ or $ \floor*{\frac{\tilde{\gamma}}{\Omega}} $ to $ \infty $, whereas the second terms' summation range takes $ \ell $ from zero to $ \floor*{\frac{\tilde{\gamma}}{\Omega}} $. The contributions to the particular solution from above the cutoff energy scale $ \tilde{\gamma} $ is highly suppressed (especially at lower PN orders) because of the properties of $ j_\ell $ and $ h^{(1)}_\ell $ for asymptotically large values of $ \ell $, and terms in $ \mathfrak{R} $ series containing the same can be approximated to zero at the current orders of calculation. Hence one can safely set the first summation in RHS of Eq. (\ref{riccipostexp}) to zero and consider only the second summation. Hence it is to be noted that the factors that depend on $ \epsilon $ in the above are only in the arguments of the Bessel function $ I_{\ell\,+\,\frac{1}{2}} $.\\
	
	Hence the only relevant factors that explicitly contain $ \epsilon $ in both the sums in Eq. (\ref{riccipostexp}), can be expanded about $ \epsilon\,\rightarrow\,0 $ as follows
	\begin{eqnarray}\label{bessprodexp}
	I_{\ell+\frac{1}{2}}\,\left(\epsilon^2\,\Lambda\,\left|X_L\right|\right) &\approx& \epsilon^{2\,\ell\,+\,1}\,\frac{\Lambda^{\frac{2\,\ell\,+\,1}{2}}}{2^{\ell\,+\,\frac{1}{2}}\,\Gamma\left(\ell\,+\,\frac{3}{2}\right)}\,\left|X_L\right|^{\ell\,+\,\frac{1}{2}}\nonumber\\
	\label{flexp}
	\end{eqnarray}
	The approximation of (\ref{bessprodexp}), when substituted in for the relevant factors of the RHS of Eq. (\ref{riccipostexp}), leads to the following polynomial series of $ \epsilon $ about $ \epsilon\,\rightarrow\,0 $
	\begin{eqnarray}\label{ricciexp}
	&&\mathfrak{R}_{part}\left(t,\,x^i\right) \approx -8\,\pi\,\gamma^2\,\sum_{L,\,\ell,\,m}^{\ell\,=\,\infty}\,\epsilon^{2\,\ell\,+\,4}\,\frac{\left(\tilde{\gamma}^2\,-\,m^2\,\Omega^2\right)^{\frac{2\,\ell\,+\,1}{4}}}{2^{\ell\,+\,\frac{1}{2}}\,\tilde{\gamma}\left(\ell\,+\,\frac{3}{2}\right)}\nonumber\\
	&&\times\,e^{i\,m\,\left(\phi_0\,-\,\Omega\,t\right)}\,\,\frac{K_{\ell\,+\,\frac{1}{2}}\left(\sqrt{\tilde{\gamma}^2\,-\,m^2\,\Omega^2}\,\left|Z^i_L\right|\right)}{\sqrt{\left|Z^i_L\right|}}\,Y_{\ell\,m}\left(Z^\theta_L,\,Z^\phi_L\right)\nonumber\\
	&&\times\,\mathfrak{M}_{L\,\ell\,m}\\
	&&\mathfrak{M}_{L\,\ell\,m} = 2\,\pi\,\left(-1\right)^m\,N_{\ell\,-m}\,\int_{\mathcal{B}_L}\left|X^i_L\right|^{2+\ell}\,\sin \left(X^\theta_L\right)\nonumber\\
	&&\times\quad P_{\ell\,-m}\left(\cos X^\theta_L\right)\,\mathfrak{T}_m\left(\left|X^i_L\right|,\,X^\theta_L\right)d\left|X^i_L\right|\,dX^\theta_L\nonumber\\
	\end{eqnarray}
	From Eq. (\ref{ricciexp}), one can immediately notice that the leading order deviation for the particular solution of the Ricci scalar for the $ L^{th} $ body zone comes at $ \mathcal{O}\left(\epsilon^4\right) $ for $ \ell\,=\,0 $ with $ \mathfrak{M}_{L\,0\,0}\,\equiv\,M_L $ being the mass monopole, and was found to be
	\begin{eqnarray}\label{ricciinhom}
	{}_{(4)}\mathfrak{R}_{part} &\equiv& {}_{(4)}\mathfrak{R} \,=\, -8\,\pi\,\gamma^2\,\epsilon^4\,\frac{M_L\,\,e^{-\frac{\gamma\,\left|Z^i_L\right|}{\sqrt{\epsilon}}}}{\left|Z^i_L\right|}
	\end{eqnarray}
	\section{Contribution of the perturbed Ricci scalar to the dynamics of $ \mathfrak{\tilde{h}}^{\mu\nu} $, and changes to the GR two-body equations of motion till 6 PN}
	\label{ap:surfint}
	The leading order particular Ricci deviation densities are given by the sum of contributions from the two body zones
	\begin{eqnarray}
	{}_{(4)}\mathfrak{R} &=& -8\,\pi\,\gamma^2\,\epsilon^4\,\sum_{L=1,\,2}\,\frac{M_L\,\,e^{-\frac{\gamma\,\left|Z^i_L\right|}{\sqrt{\epsilon}}}}{\left|Z^i_L\right|}
	\end{eqnarray}
	The new gothic tensor deviation trace at the leading order is given as follows
	\begin{eqnarray}
	{}_{(2)}\tilde{\mathfrak{h}} &\equiv& 4\,\epsilon^2\,\sum_{L\,=\,1,\,2}\,\frac{M_L}{\left|Z^i_L\right|} \label{oldpot}
	\end{eqnarray}
	The spatial derivatives of the two scalar densities are given as follows
	\begin{eqnarray}
	\partial_j\left({}_{(2)}\mathfrak{\tilde{h}}\right) &=& -4\,\epsilon^2\,\sum_{L=1,\,2}\,\frac{M_L}{\left|Z^i_L\right|^3}\,Z^L_j \label{hfrakspat}\\
	{}_{(4)}\mathfrak{R}_{\,,\,j} &=& 8\,\pi\,\gamma^2\,\epsilon^4\,\sum_{L=1,\,2}\,\frac{M_L}{\left|Z^i_L\right|^3}\,\left(1\,+\,\frac{\gamma\left|Z^i_L\right|}{\sqrt{\epsilon}}\right)\,e^{-\frac{\gamma\,\left|Z^i_L\right|}{\sqrt{\epsilon}}}\,Z_j^L\label{scalspat}
	\end{eqnarray}
	Therefore, the first appearance of the effective source term $ t^{\mu\nu}_{eff} $ in the $ \epsilon $ series of $ \tilde{\Lambda}^{\mu\nu} $, following the definition in Eq. (\ref{lambdasix}), was found to be $ \mathcal{O}\left(\epsilon^{12}\right) $ (since $ \left(-g\right)^{-2}\,\sim\,\epsilon^4 $ and $ \tilde{\mathfrak{h}}\,\sim\,\epsilon^2 $) in the following
	\begin{eqnarray}
	&&t^{i\,j}_{eff} = \frac{\epsilon^{12}}{192\,\pi\,\gamma^4}\,\left[2\,\epsilon^2\left(\eta^{i\,k}\,\eta^{j\,l}\,+\,\eta^{i\,l}\,\eta^{j\,k}\,-\,\eta^{i\,j}\,\eta^{k\,l}\right)\,{}_{(4)}\mathfrak{R}\,\partial_k{}_{(4)}\mathfrak{R}\,\partial_l\left({}_{(2)}\tilde{\mathfrak{h}}\right)\right.\nonumber\\
	&&\left.\,+\,\left(\eta^{i\,k}\,\eta^{j\,l}\,-\,\frac{1}{2}\,\eta^{i\,j}\,\eta^{k\,l}\right)\,\left\{4\,\partial_k\left({}_{(4)}\mathfrak{R}\right)\,\partial_l\left({}_{(4)}\mathfrak{R}\right)\,+\epsilon^4\,{}_{(4)}\mathfrak{R}_{hom}^2\,\partial_k\left({}_{(2)}\tilde{\mathfrak{h}}\right)\,\partial_l\left({}_{(2)}\tilde{\mathfrak{h}}\right)\right\}\right] \label{teffrelap}
	\end{eqnarray}
	
	Eq. (\ref{oldpot}) is proportional the Newtonian potential with the 'usual definition of mass', and $ \left|Z^i_L\right| $ are the radial distances from the center of mass of the two compact objects to a point in the external Minkowski space-time.\\
	
	The change in the equations of motion is given by the boundary independent part of the following surface integral, as was defined in Eq. (\ref{vevol})
	\begin{eqnarray}
	F^i_{eff} &=& -\epsilon^{-4}\,\oint_{\partial\mathcal{B}_1}\,dS_j\,t^{i\,j}_{eff}\\
	&=& -\epsilon^{-4}\,\oint_{\partial\mathcal{B}_1}\,\left|dS\right|\,r^1_j\,t^{i\,j}_{eff}\\
	&=& -\epsilon^{-4}\,\oint_{\partial\mathcal{B}_1}\,\left(\epsilon\,a_1\right)^2\,\sin\theta\,d\theta\,d\phi\,r^1_j\,t^{i\,j}_{eff}\label{surfint1}
	\end{eqnarray}
	where $ a_1 $ is the radius of the first body zone coordinate system as seen by an observer at asymptotic infinity, and $ \epsilon\,a_1 $ is the radius as seen by an observer in the first body zone coordinate system. The radius is $ \epsilon\,a_1 $ instead of just $ a_1 $ since the integral is being performed in the body zone coordinate system to account for strong internal gravity. $ r^1_k $ is a unit one-form perpendicular at the surface of the sphere that forms the boundary of the first body zone coordinate system. Substituting Eq. (\ref{hfrakspat}) and (\ref{scalspat}) in Eq. (\ref{teffrelap}), one obtains, after contracting the extra effective energy-momentum-stress density tensor $ t^{i\,j}_{eff} $ with the unit 1-form $ r_j^1 $
	\begin{eqnarray}
	&&r_j^i\,t^{i\,j}_{eff} = \frac{\epsilon^{12}}{192\,\pi\,\gamma^4}\,\left[-512\,\pi^2\,\epsilon^2\,\gamma^4\,\sum_{L_1,\,L_2,\,L_3\,=\,1}^{2}\left(Z^i_{L_2}\,r_1^l\,Z_l^{L_3}\,+\,Z^i_{L_3}\,r^l_1\,Z^{L_2}_l\,-\,r^i_1\,Z^l_{L_2}\,Z^{L_3}_l\right)\,M_{L_1}\,M_{L_2}\,M_{L_3}\right.\nonumber\\
	&&\times\,\frac{\left(1\,+\,\gamma\,\left|Z^i_{L_2}\right|\right)}{\left|Z^i_{L_1}\right|\,\left|Z^i_{L_2}\right|^3\,\left|Z^i_{L_3}\right|^3}\,e^{-\frac{\gamma}{\sqrt{\epsilon}}\,\left(\left|Z^i_{L_1}\right|\,+\,\left|Z^i_{L_2}\right|\right)}\,+\,256\,\pi^2\,\gamma^4\,\sum_{L_1,\,L_2\,=\,1}^{2}\,\left(r_1^l\,Z^i_{L_1}\,Z^{L_2}_l\,-\,\frac{1}{2}\,r^i_1\,Z^l_{L_1}\,Z^{L_2}_l\right) \nonumber\\
	&&\times\,\frac{M_{L_1}\,M_{L_2}\,\left(1+\frac{\gamma}{\sqrt{\epsilon}}\,\left|Z^i_{L_1}\right|\,\right)\,\left(1+\frac{\gamma}{\sqrt{\epsilon}}\,\left|Z^i_{L_1}\right|\,\right)}{\left|Z^i_{L_1}\right|^3\,\left|Z^i_{L_2}\right|^3}\,e^{-\frac{\gamma}{\sqrt{\epsilon}}\left(\left|Z^i_{L_1}\right|\,+\,\left|Z^i_{L_2}\right|\right)}\,+\,1024\,\pi^2\,\epsilon^4\,\gamma^4\,\sum_{L_1,\,L_2,\,L_3\,L_4\,=\,1}^{2}\nonumber\\
	&&\left.\times\,\left(r_1^l\,Z^i_{L_3}\,Z^{L_4}_l\,-\,\frac{1}{2}\,r^i_1\,Z^l_{L_3}\,Z^{L_4}_l\right)\,\frac{M_{L_1}\,M_{L_2}\,M_{L_3}\,M_{L_4}}{\left|Z^i_{L_1}\right|\,\left|Z^i_{L_2}\right|\,\left|Z^i_{L_3}\right|^3\,\left|Z^i_{L_4}\right|^3}\,e^{-\frac{\gamma}{\sqrt{\epsilon}}\left(\left|Z^i_{L_1}\right|\,+\,\left|Z^i_{L_2}\right|\right)}\right]
	\end{eqnarray}
	Contribution to the surface integral of the first two summations vanish for all combinations of $ \left\{L_1,\,L_2\right\} $ and $ \left\{L_1,\,L_2,\,L_3\right\} $, whereas the quartic product of the last term has only non-vanishing contribution to the surface integral for $ \left\{L_1,\,L_2,\,L_3,\,L_4\right\}\,=\,\left\{2,\,2,\,2,\,1\right\}\,\&\,\left\{2,\,2,\,1,\,2\right\} $, which then leads to the following
	\begin{eqnarray}\label{vectsurfint}
	&&F^i_{eff} = -\frac{32\,\pi\,\epsilon^{12}}{3}\,M_1\,M_2^3\,\lim\limits_{\epsilon\,\rightarrow\,0}\,\lim\limits_{a_1\,\rightarrow\,\infty}\oint_{\partial\mathcal{B}_1}\,\left(\epsilon\,a_1\right)^2\,\sin\theta\,d\theta\,d\phi\,\,\,\frac{e^{-\frac{2\,\gamma}{\sqrt{\epsilon}}\left|-r_{1\,2}\,n^i\,+\,\epsilon\,a_1\,r_1^i\right|}}{\left|-r_{1\,2}\,n^i\,+\,\epsilon\,a_1\,r_1^i\right|^5\,\left(\epsilon\,a_1\right)^3}\nonumber\\
	&&\times\,\left[r_1^l\,\left(-r_{1\,2}\,n^i\,+\,\epsilon\,a_1\,r^i_1\right)\,\epsilon\,a_1\,r_l^1\,-\,\frac{1}{2}\,r_1^i\,\left(-r_{1\,2}\,n^l\,+\,\epsilon\,a_1\,r^l_1\right)\,\epsilon\,a_1\,r_l^1\right]\\
	&&=\frac{32\,\pi\,\epsilon^{12}}{3}\,\frac{M_1\,M_2^3}{r_{1\,2}^4}\,e^{-\frac{2\,\gamma\,r_{1\,2}}{\sqrt{\epsilon}}}\,\oint_{\partial\mathcal{B}_1}\,\left(n^i\,-\,\frac{1}{2}\,r^i_1\,\cos\theta\right)\,\sin\theta\,d\theta\,d\phi\\
	&&=-\frac{32\,\pi\,\epsilon^{12}}{9}\,\,\frac{M_1\,M_2^3}{r_{1\,2}^4}\,e^{-\frac{2\,\gamma\,r_{1\,2}}{\sqrt{\epsilon}}}\,n^i \label{vectsurfintsimp}
	\end{eqnarray}
	which is an effect that comes at the 6 PN order in the equations of motion, which is a very small deviation from GR. In the above $ r_1^i $ is the unit normal radially outward 3-vector on the surface of the sphere $ \partial\mathcal{B}_1 $, or the body zone boundary of the first object. $ n^i $ is the unit vector pointing from the COM of the first body zone to the COM of the second body zone, whereas $ \theta $ is the angle between $ n^i $ and $ r_1^i $. About $ \epsilon\,\rightarrow\,0 $, Eq. (\ref{vectsurfint}) requires $ a_1 $ or the radius of the body zone boundary to go to infinity, as was described in \cite{Futamase2007}. Once the limit is taken, the only part of Eq. (\ref{vectsurfint}) that is independent of the body zone boundary $ a_1 $, is given by Eq. (\ref{vectsurfintsimp}).

	The above leads to the following equations of motion till 6 PN
	\begin{eqnarray}
	M_1\,\frac{dv_1^i}{ds} &=& F^i_{Newtonian} \,+\, \epsilon^2\,F^i_{1PN,GR} \,+\, \epsilon^4\,F^i_{2PN,GR} \,+\, \epsilon^5\,F^i_{2.5PN,GR} \nonumber\\
	&&\,+\,F^i_{3PN,GR}\,+\,...\,+\,\epsilon^{12}\,\left(F^i_{6PN,GR}\,-\,\frac{32\,\pi\,M_1\,M_2^3}{9\,r_{1\,2}^4}\,\,e^{-\frac{2\,\gamma\,r_{1\,2}}{\sqrt{\epsilon}}}\,n^i\right) \label{fReom1PN}
	\end{eqnarray}
	where the forms of the various PN forces till $ \mathcal{O}\left(\epsilon^6\right) $ can be found in \cite{Itoh2004}, whereas higher order dynamics can be found in \cite{Bini:2020nsb,*Bini:2020rzn,*Bini:2020wpo} and \cite{Blumlein:2020pyo,*Blumlein:2021txj}. Eq. (\ref{fReom1PN}) is an equation of motion of two classical point particles with the mutual force given by the RHS of the above. The potential is given by the the line integral of the the conservative acceleration/force-per-unit-mass along a radial direction, which in this case is along $ -n^i $ or pointing away from the COM of the first body zone (antiparallel to the displacement vector between the first and second body zone), and hence repulsive.\\
	%
	\input{MPIGH_fR2body_VerEPJC.bbl}
\end{document}

%% file: MPIGH_fR2body_VerEPJC.bbl
%